\documentclass[aps,twocolumn,showpacs,letter,superscriptaddress]{revtex4}

\usepackage{subfigure}
\usepackage{bm}
\usepackage{color}
\usepackage{graphicx}
\usepackage{epsfig}
\usepackage{rotating}
\usepackage{multirow}
\usepackage{eqnarray,amsmath,amssymb}
\usepackage{dcolumn}
\usepackage{bm}
\usepackage{subfigure}
\usepackage{bm}
\usepackage{graphicx}
\usepackage{epsfig}
\usepackage{rotating}
\usepackage{multirow}
\usepackage{dcolumn}
\usepackage{bm}

\begin{document}

\title{{Spin Hamiltonian, {Order Out of a Coulomb Phase} and Pseudo-Criticality 
in the Highly Frustrated Pyrochlore Heisenberg Antiferromagnet FeF$_{\bf 3}$}}

\author{Azam Sadeghi}
\author{Mojtaba Alaei} \email{m.alaei@cc.iut.ac.ir }
\author{Farhad Shahbazi} \email{shahbazi@cc.iut.ac.ir}
\affiliation{Department of Physics, Isfahan University of Technology, Isfahan 84156-83111, Iran}
\author{Michel J. P. Gingras}\email{gingras@uwaterloo.ca}
\affiliation{Department of Physics and Astronomy, University of Waterloo, Waterloo, ON, N2L 3G1, Canada}
\affiliation{Perimeter Institute for Theoretical Physics, 31 Caroline North, Waterloo, ON, N2L 2Y5, Canada}
\affiliation{Canadian Institute for Advanced Research, 180 Dundas Street West, Suite 1400, Toronto, ON, M5G 1Z8, Canada}

\date{\rm\today}

\begin{abstract}
FeF$_3$, with its half-filled Fe$^{3+}$ $3d$ orbital, hence zero orbital angular momentum and $S=5/2$, is often put forward
as a prototypical 
highly-frustrated classical Heisenberg pyrochlore antiferromagnet.
By employing {\it ab initio} density functional theory (DFT), 
we obtain an effective spin Hamiltonian for this material.
This Hamiltonian contains 
nearest-neighbor antiferromagnetic Heisenberg, bi-quadratic and Dzyaloshinskii-Moriya  interactions as dominant terms and
we use  Monte Carlo simulations to investigate the nonzero temperature properties of this minimal model.
We find that upon decreasing temperature, the 
system passes through a Coulomb phase, composed of short-range correlated coplanar states, 
before transforming into an ``all-in/all-out" (AIAO) state via a very weakly first order transition
at a critical temperature $T_c\approx 22$ K, in good agreement with the experimental value for a 
reasonable set of Coulomb interaction $U$ and Hund's coupling $J_{\rm H}$ describing the material.
Despite the transition being first order, 
the AIAO order parameter evolves below $T_c$ with a power-law behavior characterized by a
pseudo ``critical exponent" $\beta \approx 0.18$ in accord with experiment.
We comment on the origin of this unusual $\beta$ value.

\end{abstract}


\maketitle

Systems with magnetic moments on the vertices of two- and three-dimensional networks of corner-shared 
triangles or tetrahedra and with predominant effective antiferromagnetic nearest-neighbor (n.n.) interactions have 
tenuous tendency towards conventional long-range magnetic order \cite{Villain,springer_book}.
Consequently,  the exotic low-temperature properties of materials with such an architecture are ultimately 
dictated by the mutual competition of perturbations beyond n.n. interactions \cite{springer_book}. 


One theoretically expects such
highly-frustrated magnets to ubiquitously display a {\it Coulomb phase} (CP) 
 \cite{Henley_ARC}. This is an 
emergent state with local constraints described by a divergence-free ``spin field'' and whose defects, 
where the constraints are violated, behave as effective charges with Coulombic interactions. 
The CP and its underlying gauge theory description 
provides an elegant setting to study the effect of various perturbations \cite{Conlon} 
as well as thermal and quantum fluctuations \cite{Gingras_QSI_review}.
A telltale experimental signature of a  CP are bow-tie (``pinch points'') singularities
in the energy-integrated neutron scattering intensity pattern \cite{Henley_ARC,zinkin}.

There is good evidence that the classical spin liquid state of
spin ice materials with discrete Ising spins may be described by a 
CP \cite{Henley_ARC,Castelnovo_ARC,Fennell_science, Morris_science}.
Unfortunately, there are few, if any, materials with continuous symmetry spins 
 that display a CP, as may be signalled by pinch points \cite{zinkin}.
For example, in Y$_2$Mo$_2$O$_7$, complex orbital effects \cite{Motome,Silverstein} 
and spin glass behavior \cite{gingras_YMO,gardner_YMO}  irradicate the CP. 
In the ZnCr$_2$O$_4$ spinel,   pinch points are not observed \cite{Lee_pinch},  
 likely because perturbations beyond  n.n. interactions and spin-lattice coupling 
eliminate them already at high temperature in the paramagnetic state \cite{Conlon}.
In this letter we propose that 
FeF$_3$,  with magnetic Fe$^{3+}$ ions on a pyrochlore network of corner-sharing tetrahedra, may be 
a strong contender for a CP with Heisenberg spins.

\begin{figure}[ht]
\centering
\includegraphics[width=0.45\columnwidth]{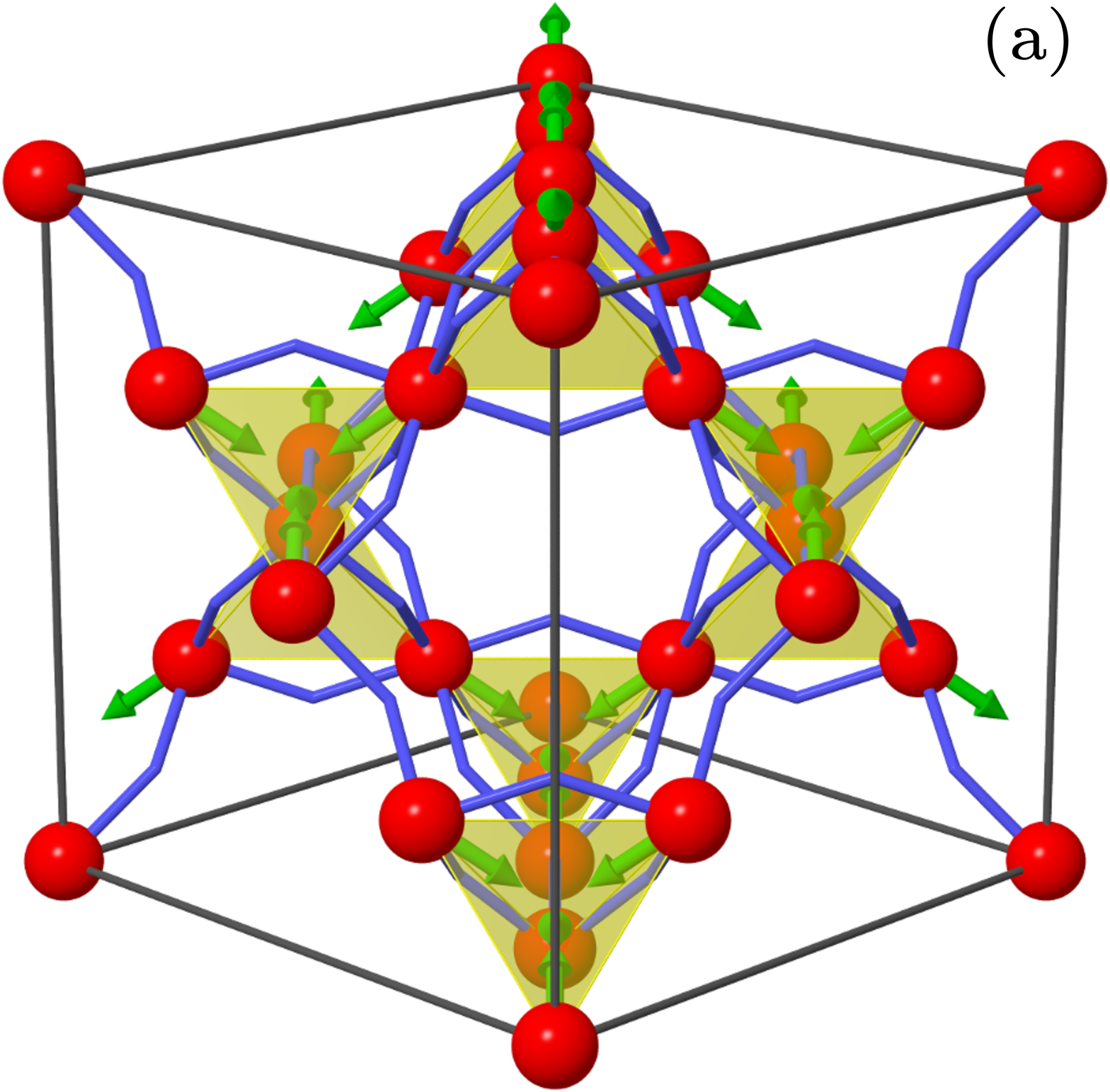}
\includegraphics[width=0.45\columnwidth]{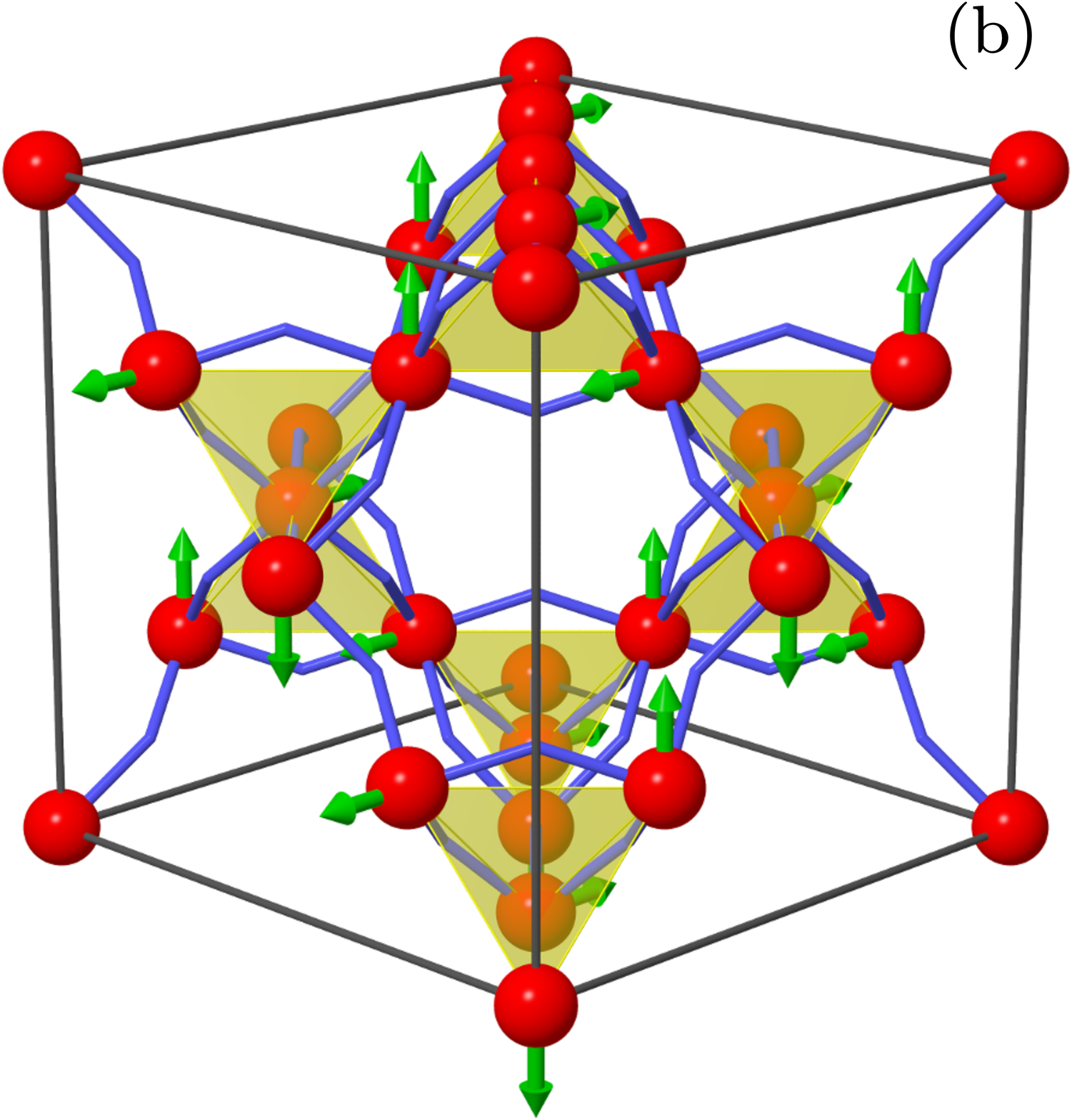}
\caption{(Color online): The structure of FeF$_3$. 
Red (dark grey) spheres denote the Fe$^{3+}$ ions with their spin indicated by a green arrow. 
The  F$^{-} $ ions (not shown) are located at the (shown) 
bents where bonds merge.
(a) The AIAO state. (b) A coplanar spin configuration (for clarity, a long-range coplanar state is shown).
}
\label{fig:FeF3}
\end{figure}

With Fe$^{3+}$ being a $3d$ S-state (spin-only) $S=5/2$ ion, 
single-ion anisotropy and anisotropic spin-spin interactions should be small in FeF$_3$,
making it a good candidate material with predominant  n.n. Heisenberg exchange.
Neutron scattering and M{\"o}ssbauer 
experiments find  long-range magnetic order below $T_c \approx 20^{+2}_{-5}$ K \cite{ferey86-1,ferey86-2,ferey87,reimers1,reimers2}.  
Yet, the  static magnetic susceptibility shows a deviation from the 
Curie-Weiss law even at $300$ K, implying the existence of
strong antiferromagnetic exchange and short-range correlations extending up to temperatures much higher than $T_c$ \cite{ferey86-1} and thus a very 
high degree of frustration \cite{springer_book,Ramirez_frust}.
The ordered phase is an ``all-in/all-out''  (AIAO) state   \cite{ferey86-1}
in which the Fe$^{3+}$ magnetic moments point from the corners to the centers (or vice versa) of each tetrahedron  (see Fig.~\ref{fig:FeF3}a).
Notably,  neutron diffraction experiments find a power-law growth of the AIAO order parameter  characterized by a ``critical 
exponent'' $\beta \sim 0.18$ \cite{reimers2}. 
This value  differs significantly from standard order-parameter exponents $\beta \sim 1/3$ for three-dimensional systems,
which prompted the suggestion of an underlying ``new'' universality class~\cite{reimers2}. 
There appears to have been no attempt to 
determine a realistic 
spin Hamiltonian ${\cal H}$ for FeF$_3$.
In this paper, we  employ density-functional theory (DFT) to flesh out such ${\cal H}$  and use it
to study the development of correlations upon
approaching $T_c$ and to explore the associated critical properties.
By computing the energy of various spin configurations and performing 
Monte Carlo simulations, we expose  
a highly entropic coplanar (Coulombic) state above $T_c$ and its demise at $T\le T_c$ 
against an energetically selected AIAO state along with replicating the 
unusual  $\beta \sim 0.18$ exponent.


{\it Spin Hamiltonian and DFT calculations} $-$ 
 The classical spin Hamiltonian for FeF$_3$ is given by
\begin{equation}
\label{H.eq}
{\cal H}={\cal H}_{\rm {H}}+{\cal H}_{\rm{b.q.}}+{\cal H}_{\rm{r}}+{\cal H}_{\rm{DM}}+{\cal H}_{\rm{s.i.}}.
\end{equation}
${\cal H}_{\rm{H}}=\sum_{i>j} J_{ij} \, {\bm S}_i \cdotp {\bm S}_j$ denotes the isotropic Heisenberg term.
${\bm S}_i$ and ${\bm S}_j$ are classical unit vectors representing the orientation of the magnetic moments 
at sites $i$ and $j$, respectively.
We consider a distance-dependent exchange $J_{ij}$ between ${\bm S}_i$ and ${\bm S}_j$, with first ($J_1$),
second ($J_2$) and two distinct third ($J_{3a}$ and $J_{3b}$) n.n. \cite{J3}.
 ${\cal H}_{\rm{b.q.}}=\sum_{i>j} B_{ij} \,  ({\bm S}_i \cdotp {\bm S}_j)^2$ 
is the bi-quadratic interaction with n.n. coupling $B_1$.
$ {\cal H}_{\rm{r}}=\sum_{ijkl}  K
[ ({\bm S}_i \cdotp {\bm S}_j)({\bm S}_k \cdotp {\bm S}_l) 
+({\bm S}_j\cdotp {\bm S}_k)({\bm S}_l \cdotp {\bm S}_i) 
-({\bm S}_i \cdotp {\bm S}_k)({\bm S}_j \cdotp {\bm S}_l)]$ is the ring-exchange interaction.
The last two (anisotropic interaction) terms, originating from  spin-orbit coupling (SOC),  
 are the Dzyaloshinskii-Moriya (DM) interaction, 
$ {\cal H}_{\rm{DM}}=D \sum_{\langle i,j \rangle} 
{\hat{\bf D}}_{ij} \cdotp ( {\bm S}_i \times {\bm S}_{j})$,  and  single-ion anisotropy
$ {\cal H}_{\rm{s.i.}}=	\Delta \sum_{i} ({\bm S}_{i}\cdot {\hat{\bf d}_{i}})^{2}$.
${\hat{\bf D}}_{ij}$  are the DM (unit) vectors determined according to the Moriya rules \cite{elhajal,supplement}.
The unit vector ${\hat{\bf d}}_{i}$ denotes the 
single-ion easy-axis along the local cubic $[111]$ direction  at site $i$. 

We next use DFT to study the properties of FeF$_3$.
For all  computations, the experimental data for the conventional cubic unit cell 
lattice parameter (10.325 \AA{}) and position of the ions  were used~\cite{ferey86-2}. 
The DFT calculations were carried out with the full-potential 
linearized augmented plane wave (FLAPW) method, employing  the Fleur code~\cite{fleur}. 
We used the local density approximation (LDA) to  account for the  electron exchange-correlation.
Electron-electron interactions due to the on-site electron repulsion $U$ are taken into account  
 using the LDA+$U$ method.  
The effective on-site Coulomb interaction, $U_{\rm eff}$, is defined as  
$U_{\rm eff}=U-J_{\rm H}$, where $U$ is the bare Coulomb repulsion and $J_{\rm H}$ is the on-site 
ferromagnetic Hund's exchange, which we set to 1.0 eV,
a typical value in such DFT calculations. 
Using a linear response approach~\cite{lda+u}, we obtain $U_{\rm eff} \approx 2.8$ eV from the 
Quantum-Espresso code~\cite{QE}.
The influence of $U_{\rm eff}$ on various properties is discussed in the Supplementary Material (S.M.)\cite{supplement}. 
The mininimum energy states possess a global continuous $O(3)$ degeneracy within LDA+$U$.
However, incorporating the effect of SOC within LDA+$U$+SOC 
leads to an AIAO configuration with spins along $\langle 111 \rangle$ as minimum energy state. 
We find FeF$_3$ to be an insulator with a $1.04$ eV band gap within LDA+SOC.
 The band gap rises to $2.49$ eV in LDA+$U$+SOC  with  $U_{\rm eff} = 2.8$ eV.

We next determine the coupling constants of ${\cal H}$ 
 using spin-polarized DFT calculations.
For the first three (isotropic) terms of Eq.~(\ref{H.eq}), 
we use LDA+$U$ to compute the total energy  difference  between
various  magnetic configurations \cite{supplement}. 
We assume that $J_{3b}$ \cite{J3} as well as 
farther Heisenberg exchanges ($J_{m},m \ge 4 $), and bi-quadratic terms further than first n.n.
($B_m,m \ge 2 $) are negligible. 
By  matching the energy differences for spin-polarized electronic states
with that of ${\cal H}$, we determine $J_1, J_{2}, J_{3a}$ and $B_1$ \cite{supplement}. 
To compute the anisotropic DM ($D$) and single-ion ($\Delta$) couplings  
arising from SOC, we use the LDA+$U$+SOC framework.
We consider non-collinear spin-polarized configurations,
keeping the isotropic terms of ${\cal H}$ unchanged \cite{supplement}.  
The largest couplings within LDA+$U$+SOC are (all in meV):
\begin{eqnarray}
J_1 =  32.7,   \; J_2=0.6,	\;	J_{3a} = 0.5, \; B_1 =  1.0,	 \;	D=0.6 .
\label{eq:couplings}
\end{eqnarray}
The ring-exchange $K$ and the single-ion coupling 
$\Delta $ are found to be smaller than 0.1 meV \cite{supplement},  
so we henceforth ignore them.
The Curie-Weiss temperature, $\theta_{\rm CW}$, can thus be estimated by
 $\theta_{\rm CW} \sim qJ_1/3 \sim 760$ K, where $q=6$ is the number of n.n.
With $\theta_{\rm CW}/T_c \sim 38$, we thus confirm FeF$_3$ to be  a highly-frustrated
antiferromagnet \cite{springer_book,Ramirez_frust}.


{\it Ground states and Monte Carlo simulations} $ - $
Following Refs.~[\onlinecite{enjalran,reimers3}], 
 we find that mean-field theory predicts 
 AIAO order  for ${\cal H}$ with the above 
 $\{J_1,J_2,J_{3a},D\}$ values and $B_1 \equiv 0$. 
This is confirmed by MC simulations when including $B_1=1.0$ 
meV since $(B_1>0,D=0)$ stabilizes an $O(3)$ symmetric AIAO state 
(see discussion below). 
In the rest of the paper, we focus on the  {\it generic} 
aspects of the collective behavior of the system (such as exponent $\beta \sim 0.18$).
 While specific details ($T_c$, and the Fe and F magnetic moments)
 depend on the value of the ($U,J_{\rm H}$) parameters \cite{supplement},
  we expect the overall collective properties  to survive small adjustments 
of these parameters \cite{supplement}.  Therefore, to explore those generic facets, 
we consider a minimal model Hamiltonian, ${\cal H}_{\rm min}$, 
with  ${\cal H}_{\rm min} \equiv {\cal H}(J_1,B_1,D,J_2=J_{3a}=0)$ with the 
$(J_1,B_1,D)$ values of Eq.~(\ref{eq:couplings}).

The ground state of ${\cal H}_{\rm min}$ with ($J_1>0,B_1=D=0$), 
is highly degenerate on the pyrochlore lattice \cite{Villain,reimers3,moessner1,moessner2}. 
The ground state manifold consists of spin configurations with vanishing total spin on each tetrahedron, 
 with two continuous degrees of freedom per tetrahedron \cite{supplement,reimers3,moessner1,moessner2}.  
The minimum energy of ${\cal H}_{\rm min}$ with ($J_1>0,B_1>0,D=0$)
has a globally $O(3)$ degenerate non-coplanar AIAO spin configuration with an angle $109.47^\circ$ between 
each n.n. pair of spins \cite{supplement}.
 Including $D>0$ fixes the spin directions within such a configuration to one of two {\it discrete} 
AIAO states with spins
along the cubic $\langle 111\rangle$ directions \cite{supplement}.
With $B_1=0$, direct DM interactions ($D> 0$)  also dictates an AIAO state \cite{elhajal}.
The ground state energy per spin \cite{supplement} for the coplanar and AIAO state is,  respectively, 
$\epsilon_{\rm{coplanar}} = -J_{1} + B_1   - \sqrt{2}D$ 
and $\epsilon_{\rm{AIAO}} =  -J_{1} + B_1/3 -2\sqrt{2}D$,
showing that the ground state is AIAO {\it for all} $B_1>0$ 
and $D>0$ values.

With ($J_1>0,B_1>0,D=0$),  ${\cal H}_{\rm min}$ 
displays for a tetrahedron three saddle points in its energy landscape which correspond
 to coplanar states \cite{supplement}. 
 In these   states,  two out of four spins are antiparallel along a given axis and
perpendicular to the other axis along which the two remaining spins are themselves  aligned mutually antiparallel.
The addition of $D> 0$  restricts the orientation of the ``coplanes'' to be along the
$ xz $, $ xy $   or $ yz $ planes of the cubic unit cell, 
depending on which  pairs of spins are chosen to be collinear~\cite{supplement}.   
There is an exponentially large number of such coplanar states which 
provide an entropy buffer above the critical temperature
where the system orders into AIAO.
One such coplanar spin arrangement,
within the $xz$ plane, is depicted in Fig.~\ref{fig:FeF3}b. 

\begin{figure}[t]
\includegraphics[width=.9\columnwidth]{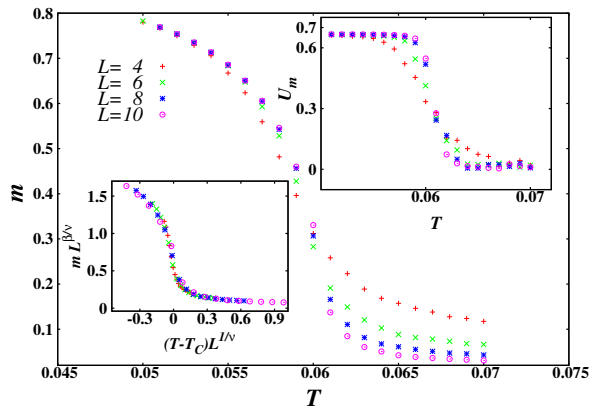}
\caption{(Color online) main panel: variation of the AIAO order parameter ($m$) versus 
temperature (in units of $J_1$), for lattices of linear size $ L=4,6,8,10 $. 
Top inset: Fourth order Binder cumulant of $m$ versus temperature, $T$ (in units of $J_1$),
for the same lattice sizes. 
Left inset: Finite size scaling of $m(t,L)$ 
 with $\beta=0.18(2)$ and $\nu=0.60(2)$.    }
\label{M}
\end{figure}

We next perform  Monte Carlo simulations to gain some insight into the
 finite temperature  properties of ${\cal H}_{\rm min}$.
 We use standard single-spin Metropolis algorithm on lattices consisting of 
 $N=4\times L^3$  spins, where $L$ is the linear dimension of the rhombohedral simulation cell. 
To ensure thermal equilibrium, $10^6$ 
Monte Carlo steps (MCS) per spin were used for each temperature and 
 $10^6$ MCS for the data collection. 
To reduce the correlation between measurements, $10$  to $20$ MC sweeps 
were discarded between successive data collection. 
To ascertain that our results are fully thermally equilibrated and are 
not caused by a two-phase coexistence, we started the simulation runs  from diffferent initial
states, i.e totally disordered, AIAO ordered and
coplanar states and checked that all final results remain the same.


Quantities of particular interest are the  AIAO order  parameter 
$m \equiv  {\Sigma_{i,a} {\bm S}_i^a \cdot {\hat{\bf d}}^a}/{N}$ 
(${\hat{\bf d}}^a$ is the local cubic $[111]$ direction for sublattice $a$)
 and the Binder fourth order cumulant for both $m$ and energy $E$, 
defined respectively as     
$U_m(T)	\equiv 1-\frac{1}{3}\frac{\langle m^4 \rangle}	{{\langle m^2 \rangle}^2}$ and 
$U_{E}(T) \equiv  1-\frac{1}{3}\frac{\langle E^4 \rangle  }  {{\langle E^2 \rangle}^2}$.
$U_m$   vanishes in the paramagnetic phase, with a Gaussian probability distribution  for $m$,
while $U_m$ approaches $2/3$ in the ordered phase~\cite{binder1,binder2,binder3}. 
$U_E$ tends asymptotically to $2/3$ in both 
the ordered and paramagnetic phase while reaching a minimum, $U_E^{\rm min}$,
near the transition \cite{supplement}. 



 \begin{figure*}[t]
\centering
\includegraphics[width=2.\columnwidth, height=0.5\columnwidth]{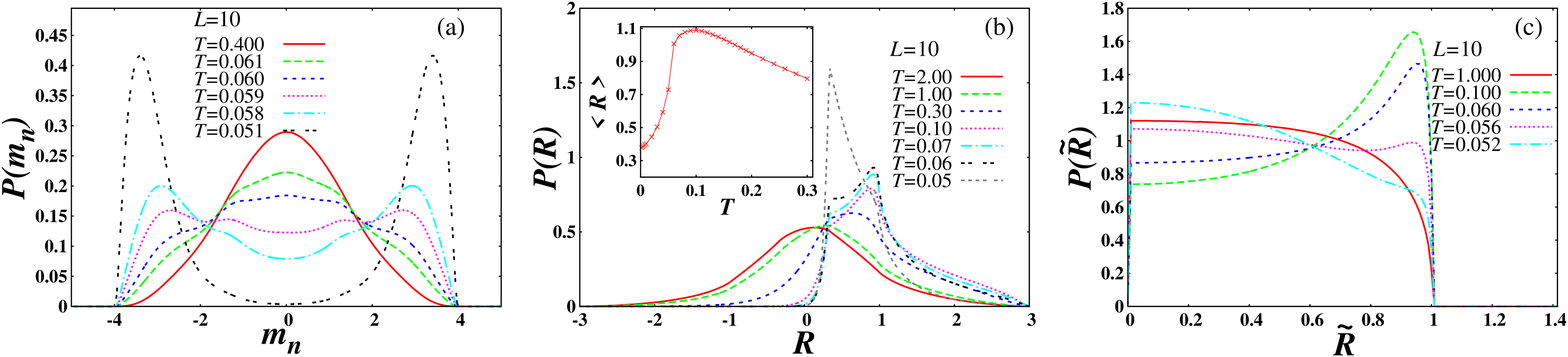}
\includegraphics[width=2.\columnwidth, height=0.5\columnwidth]{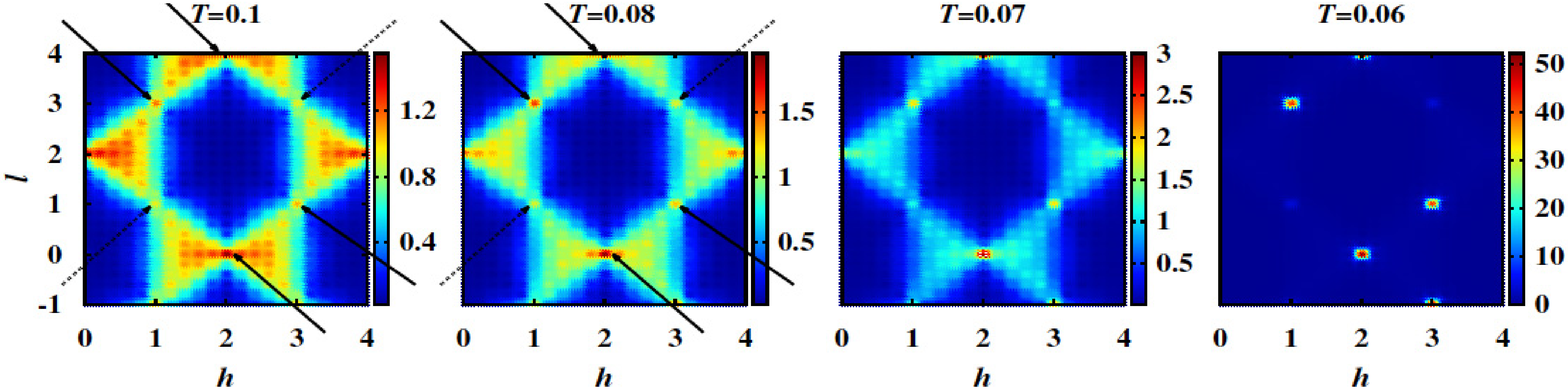}
\caption{(Color online) {\bf Top row:} 
Probability distribution functions, $P(m_n)$, $P(R)$ and $P({\tilde R})$, 
 as a function of temperature $T$ and for a lattice of linear size $L=10$. 
The inset of panel (b) shows the $T$ dependence of $\langle R\rangle$, 
which displays a sharp drop at $T_c\approx 0.06$,  a further
 indication for the discontinuous nature of the transition.    
{\bf Bottom row:} Temperature evolution of the 
neutron structure factor, $S({\bm q})$, in the $(hhl)$ plane as $T$ approaches $T_c$ from
the paramagnetic phase. The arrows indicate the location of pinch points for the
$T=0.1$ and $T=0.08$ panels (see text).
}
\label{PDF}
\end{figure*}

The temperature dependence of $m$ and $U_m$ is shown in the  main panel and top inset of Fig.~(\ref{M}).
Both plots indicate a narrow critical region around $T \approx 0.06$. 
The left inset in Fig.~\ref{M} shows the finite-size scaling of $m$ for different $L$ 
according to the finite-size scaling behavior
${m }  =  L^{-\beta / \nu} {\cal M}(t L^{1/\nu})$.
Here $t \equiv  (T_c-T)/T_c$ is the reduced temperature, $\beta$ is the order parameter exponent, 
 $\nu$ is  the correlation length exponent and ${\cal M}$ is the scaling function \cite{binder3}.
 This  analysis yields   $T_{c}/J_1=0.0601(2)$, $\beta=0.18(2)$ and $\nu=0.60(2)$. 
With $J_1=32.7$ meV = $379.47$ K, we get $T_c\approx 22$ K, in good  agreement with the   
experimental value~\cite{ferey86-1,ferey86-2,ferey87,reimers2}.
Perhaps most noteworthy,  the Monte Carlo  exponent $\beta \approx 0.18$ value 
corresponds to that found in experiment \cite{reimers2}.
 While these scaling arguments naively suggest that the transition is second order,
it is instructive to consider the  $L$ dependence of $U_E^{\rm min}$ which,
for a first order transition, is given by \cite{binder3}, 
$U_{E}^{\rm min}(L)=U^*+AL^{-d}+{\cal O}(L^{-2d})$ ,
with  $U^* < 2/3$. Here $d=3$ is the space dimension and $A$ is a constant.
The precise linear fit of  $U^{\rm min}_{E}(L)$) versus  $L^{-3}$, 
with $U^*=0.666664(1)$, hence very close to $2/3$, 
that we find (see Fig.~\ref{E-binder} in the S.M. \cite{supplement}) 
suggests that the transition might actually be very weakly first order.



To shed further  light on the nature of the transition, 
we compute the probability distribution function of the order parameter 
per tetrahedron, $P(m_n)$, with $m_n \equiv \Sigma_{a=1}^{4} {\bm S}^a \cdot 
{\hat{\bf d}}^a$. We also compute the  probability distribution function of 
two distinct four-spin correlations within each tetrahedron, $P(R)$ and $P({\tilde R})$, with 
\footnotesize
\begin{equation}
\begin{split}
\label{R}
&{R \equiv ({\bm S}_{1}\cdot{\bm  S}_{2})({\bm  S}_{3}\cdot{\bm  S}_{4})
+ ({\bm  S}_{1}\cdot{\bm  S}_{3})({\bm  S}_{2}\cdot{\bm  S}_{4})
+ ({\bm  S}_{1}\cdot{\bm  S}_{4})({\bm  S}_{2}\cdot{\bm  S}_{3})},\nonumber\\
&{\tilde R} \equiv |({\bm S}_{1}\cdot{\bm  S}_{2})({\bm  S}_{3}\cdot{\bm  S}_{4})
-({\bm  S}_{1}\cdot{\bm  S}_{3})({\bm  S}_{2}\cdot{\bm  S}_{4})
+ ({\bm  S}_{1}\cdot{\bm  S}_{4})({\bm  S}_{2}\cdot{\bm  S}_{3})|.\nonumber
\end{split}
\end{equation}  
\normalsize
Figures \ref{PDF}a, \ref{PDF}b and \ref{PDF}c show  $P(m_n)$, $P(R)$ and $P({\tilde R})$
versus $T$ for $L=10$.  
$P(m_n)$ is a Gaussian centered at $m_{n}=0$ for $T\gg T_c$.
As $T$ decreases,  $P(m_n)$  deviates from a Gaussian near $T_c$, 
developing {\it four}  peaks with $m_n\neq 0$   for $T\lesssim T_c$.
Well  below the transition, only  two peaks   at  $|m_n| \approx 4$ remain,  
corresponding  to almost perfect AIAO order.
The peculiar temperature evolution of $P(m_n)$ suggests 
that another state coexists or competes with the AIAO state near $T_c$.
The nature of this other state can be  clarified  by considering $P(R)$ and $P({\tilde R})$ in Figs. \ref{PDF}b and \ref{PDF}c, respectively.
Two peaks arise in $P(R)$ at $T\gtrsim T_c$; one at $R \approx 1/3$ and another at $R \approx 1$ 
(see panel \ref{PDF}b).
The former corresponds to an AIAO spin configuration for which 
$({\bm  S}_{a}\cdot{\bm S}_{b})=-\frac{1}{3}$  at $T\ll J_1$ for two n.n. spins.
The peak at $R=1$ is consistent with coplanar states as deduced from Eq.~(\ref{R}).
Considering $P({\tilde R})$ in Fig.~\ref{PDF}c, one observes a peak at ${\tilde R} \approx 1$ near $T_c$.
One can easily show \cite{supplement} that the two equations for $R=1/3$ and ${\tilde R}=1$ have 
\emph{no common} solution for a zero net spin/moment on a tetrahedron.
Therefore, an AIAO state does not produce the peak at ${\tilde R} \approx 1$, which must therefore 
originate from the competing state.
One can show that Eqs.~(\ref{R}) for $R=1$ and ${\tilde R}=1$ admit three solutions \cite{supplement}, which are 
precisely the $xy$, $xz$ and $yz$ coplanar states discussed above.
The ``competing state'' at $T\gtrsim T_c$ is therefore short-range coplanar,
as illustrated further in the S.M. \cite{supplement},
is divergence-free in the ``spin field'' and should thus be viewed as  a CP \cite{Henley_ARC}. 
To expose further the CP nature of the state at $T\gtrsim T_c$, we compute the neutron structure factor
$S({\bm q})$ (second row of Fig. \ref{PDF}) in the $(hhl)$ scattering plane as a function of $T$.
At $T=0.1$, clear pinch points (marked by arrows) are visible. 
Some of these pinch points (solid arrows) turn into magnetic Bragg preaks ($T\sim 0.06$) while
others (dashed arrows) become mere weak diffuse spots (forbiden Bragg peaks \cite{supplement})
upon going through the transition to AIAO order at $T_c$ (see $T=0.08$, $T=0.07$ and $T=0.06$ panels in
bottom row of Fig. \ref{PDF}).

{\it Conclusion} $ - $ Using DFT, we determined the predominant couplings of the spin Hamiltonian
of the FeF$_3$ pyrochlore  Heisenberg  antiferromagnet.
We find that bi-quadratic exchange and anisotropic direct Dzyaloshinskii-Moriya interactions
conspire to select an all-in/all-out ground state. Monte Carlo simulations find a transition 
to that state at a critical temperature $T_c\approx 22$ K, in good agreement with experiments.
The transition is characterized by an order parameter pseudo ``critical exponent'' $\beta\approx 0.18$,
that is also in agreement with experiment. We view this exponent not as signalling an unusual universality class, but
rather as an effective power-law parametrization near a very weakly first order transition, perhaps near 
a mean-field tricritical point for which $\beta=1/4$ (up to logarithmic correction because three-dimensions is
the upper critical dimension for tricritical behavior \cite{stephen}).
Indeed, for $D/J_1 \lesssim 0.01$, the transition is
found to be strongly first order while it is 
second order and in the three-dimensional Ising universality class for $D/J_1 \gtrsim 0.1$ \cite{kawaholds}.
We find the state above $T_c$ to be composed of entropically favored coplanar states without long-range magnetic order
and thus a Coulomb phase \cite{Henley_ARC}.
We hope that our study will motivate a new generation of experiments on  FeF$_3$,  
perhaps even on single-crystal samples, which we would anticipate on the basis of our  work
to display interesting properties heretofore unexposed in highly frustrated Heisenberg pyrochlore antiferromagnets.

\begin{acknowledgments}

We thank Bob Cava, Peter Holdsworth, Takashi Imai, Hikaru Kawamura, Seunghun Lee, Paul McClarty, Nic Shannon, Oleg Tchernyshyov 
and Anson Wong for useful discussions. We acknowledge Hojjat Gholizadeh for help with the pyrochlore lattice figures.
This work was made possible by the facilities of the Shared Hierarchical 
Academic Research Computing Network (SHARCNET:www.sharcnet.ca) and Compute/Calcul Canada.
One of us (MG)  thanks Harald Jeschke for most useful discussions regarding DFT calculations for magnetic systems.
MG acknowledges support from the Canada Council for the Arts and the Perimeter Institute for Theoretical Physics. 
Research at PI is supported by the Government of Canada through Industry Canada and by the 
Province of Ontario through the Ministry of Economic Development \& Innovation.

\end{acknowledgments}

\section*{Supplemental  Material}

\title{Supplementary Material for: Spin Hamiltonian,  Order Out of a Coulomb Phase and Pseudo-Criticality 
in the  Frustrated Pyrochlore Heisenberg Antiferromagnet FeF$_{\bf 3}$}

\author{Azam Sadeghi}
\author{Mojtaba Alaei} 
\author{Farhad Shahbazi} 
\affiliation{Department of Physics, Isfahan University of Technology, Isfahan 84156-83111, Iran}
\author{Michel J. P. Gingras} 
\affiliation{Department of Physics and Astronomy, University of Waterloo, Waterloo, ON, N2L 3G1, Canada}
\affiliation{Perimeter Institute for Theoretical Physics, 31 Caroline North, Waterloo, ON, N2L 2Y5, Canada}
\affiliation{Canadian Institute for Advanced Research, 180 Dundas Street West, Suite 1400, Toronto, ON, M5G 1Z8, Canada}





\maketitle

In this Supplemental Material, we present details to assist the reader with the main part of the paper.
Section \ref{sec:DFT} provides detailed information as to how the interaction parameters of the spin Hamiltonian,
 ${\cal H}$ in Eq. (1) in the paper,
were determined from the density functional theory (DFT) calculations.
{{Section \ref{sec:QF} briefly discusses the question of quantum fluctuations of the Fe moments in FeF$_3$ and also the dependence of the 
magnetic moments of the Fe and F ions on the choice of the muffin-tin radius.
Section \ref{sec:DFT_dep} explores how the properties of the system depend on 
the effective Coulomb interaction, $U_{\rm eff}$. 
Section \ref{sec:biquad} discusses the nature of the energy landscape of the bi-quadratic part of the spin Hamiltonian and how it displays a
saddle point.
Section \ref{sec:DM} describes the orientation of the Dzyaloshinskii-Moriya (DM) vectors.
Section \ref{sec:BinderUE} shows the finite-size evolution of the energy Binder cumulant, $U_E$, referred to in the main text.}
In Section \ref{coplanar}, we present a complementary proof to the one presented in the main text in favor of the existence  coplanar states above the transition point.
Section \ref{Sq} discusses the details of the neutron structure function calculations reported in the main text.

\section{Ab initio derivation  of the  spin Hamiltonian for the pyrochlore-$\mbox{FeF}_3$ }
\label{sec:DFT}

To derive the spin Hamiltonian, we  compute  the energy difference between some chosen  magnetic configurations using the LDA and LDA+$U$ methods.
In the following subsections, we illustrate the method for the derivation of 
the isotropic, i.e. Heisenberg,  4-spin ring  and bi-quadratic  exchanges, as well as the anisotropic terms such as the DM and the single-ion interactions.

\subsection{Technical Details}
To compute the  Heisenberg exchange couplings,  we use a super-cell containing  $16$ Fe and $48$  F atoms (see Fig. \ref{fig:spins}), 
while  for the 4-spin ring, bi-quadratic, single-ion and the 
DM coupling constants, we use the primitive cell of pyrochlore-$\mbox{FeF}_3$, which contains 4 Fe atoms.

A muffin-tin radius of $2.2$ (au) and $1.35$ (au) is used for the  Fe and F ions, 
respectively. The cut-off  wave-vector $k_{\rm max}$= $3.8 \, {\rm au}^{-1}$ is taken for the expansion of the wave function in the interstitial region
and 64 $k$-points are picked up for performing the Brillouin zone integration. 
Although the magnetic moment of F is much less than Fe (see Section  \ref{sec:QF}),
 we find that choosing the direction of the fluorine  magnetic moments
plays a crucial role in the minimization of the total energy. Our calculations show that 
for a fixed   direction of the Fe magnetic moments,   
the direction of a given  fluorine moment in a Fe-F-Fe bond  is uniquely determined  
 as $\widehat{\mu}_{\rm F}\parallel (\widehat{\mu}_{\rm{Fe}_1} + \widehat{\mu}_{\rm{Fe}_2})$, 
where $\widehat{\mu}_{\rm{Fe}_1}$, $\widehat{\mu}_{\rm{Fe}_2}$  are the magnetic moment directors of the two neighboring   Fe ions.  
As an  example, choosing  a direction perpendicular to the optimized direction,   increases the total energy by  about 14 meV per 
Fe ion. 

\subsection{Spin Hamiltonian parameters}

\subsubsection{4-spin ring exchange}

We begin with the calculation of the 4-spin ring-exchange ($K$). To proceed, we compare the energy
 of three collinear configurations within each tetrahedron. 
We use $n_i=\pm 1$ to indicate the  direction of the magnetic moment of the ion at site $i$, along an arbitrary direction (say the $z$-axis). 
The  configurations are chosen as 
$C_1 \equiv \{n_1=n_2=n_3=n_4=1\}, C_2 \equiv \{n_1=n_2=1,n_3=n_4=-1\}, C_3 \equiv \{n_1=-1,n_2=n_3=n_4=1\} $.
The total energy differences (per primitive cell) between these configurations  are  
\begin{eqnarray}
\label{K}
E_1-E_2&=&16J_1+32J_2  \nonumber\\ 
E_1-E_3&=& 12J_1+24J_2+9K,
\end{eqnarray}
where $E_1$, $E_2$ and $E_3$ is the total energy of $C_1$, $C_2$ and $C_3$ configuration, respectively. 
These equations yield
\begin{equation}
(E_1-E_2)-\frac{4}{3} (E_1-E_3)= -12K.
\end{equation}
Our calculation within LDA gives  $K\sim 0.1$ meV. 
However, LDA+$U$ with $U_{\rm eff}=2.8$ eV gives an even  smaller  value for $K$.
We therefore ignore the 4-spin ring-exchange term in the rest  of our calculations.

\subsubsection{Heisenberg exchange parameters }

\begin{figure}
    \includegraphics[width=\columnwidth]{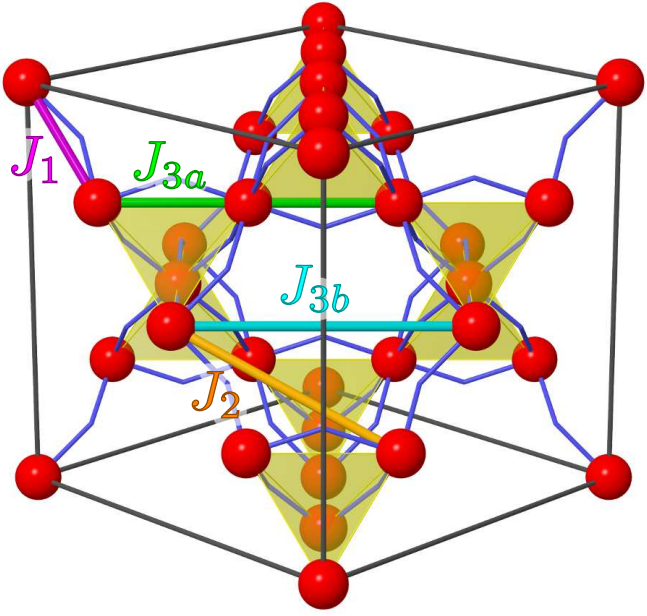}
    \caption{The Heisenberg interactions  $J_1$, $J_2$, $J_{3a}$ and $J_{3b}$ in pyrochlore FeF$_3$. 
The red spheres represent the Fe$^{3+}$ ions. The  F$^{-} $ ions (not shown) are located at the bent crossings of the Fe-F-Fe (purple) bonds. } 
    \label{fig:Js} 
\end{figure}

The site-connectivity for the  Heisenberg couplings  ${J}_{1}$, 
${J}_{2}$,   $J_{3a}$ and $J_{3b}$ is illustrated in Fig.~\ref{fig:Js}.  
Taking the  collinear spin configurations, A, B, C and D illustrated in Fig.~\ref{fig:spins}, 
one can show that the bi-quadratic term does not affect the energy differences,  
and therefore the only contributions to the total energy differences between these configurations come solely from  the Heisenberg terms. 
Invoking the spin configurations in  Fig.~\ref{fig:spins}, we
obtain the  following expressions for the total energy (per super-cell) of each configuration:

\begin{eqnarray}
\label{diff}
E_{\rm A} &=& 48 J_1 + 96 J_2 + 48 J_{3,a} + 48 J_{3, b} \\ \nonumber
E_{\rm B} &=& 24 J_1 \\ \nonumber
E_{\rm C} &=& 48 J_{3,a} + 48 J_{3,b} \\ \nonumber
E_{\rm D}&=& 12 J_1 - 16 J_2 -8 J_{3,a} - 8 J_{3, b}. 
\end{eqnarray}

LDA+$U$ calculations with $U_{\rm eff}=2.8$ eV   result  in the energy differences  
$E_{\rm A}-E_{\rm B}=866.4$ meV,  $E_{\rm A}-E_{\rm C}=1627.9$ meV, and $E_{\rm A}-E_{\rm D}=1274.9$ meV, 
from which,  assuming $J_{3,b} \ll J_{3,a}$,  we get 
 $J_1=32.7$ meV, $J_2=0.6$ meV and $J_{3,a}=0.5$ meV.

\begin{figure}
\centering
\begin{tabular}{cc}
A &
B\\
\includegraphics[width=0.5\columnwidth]{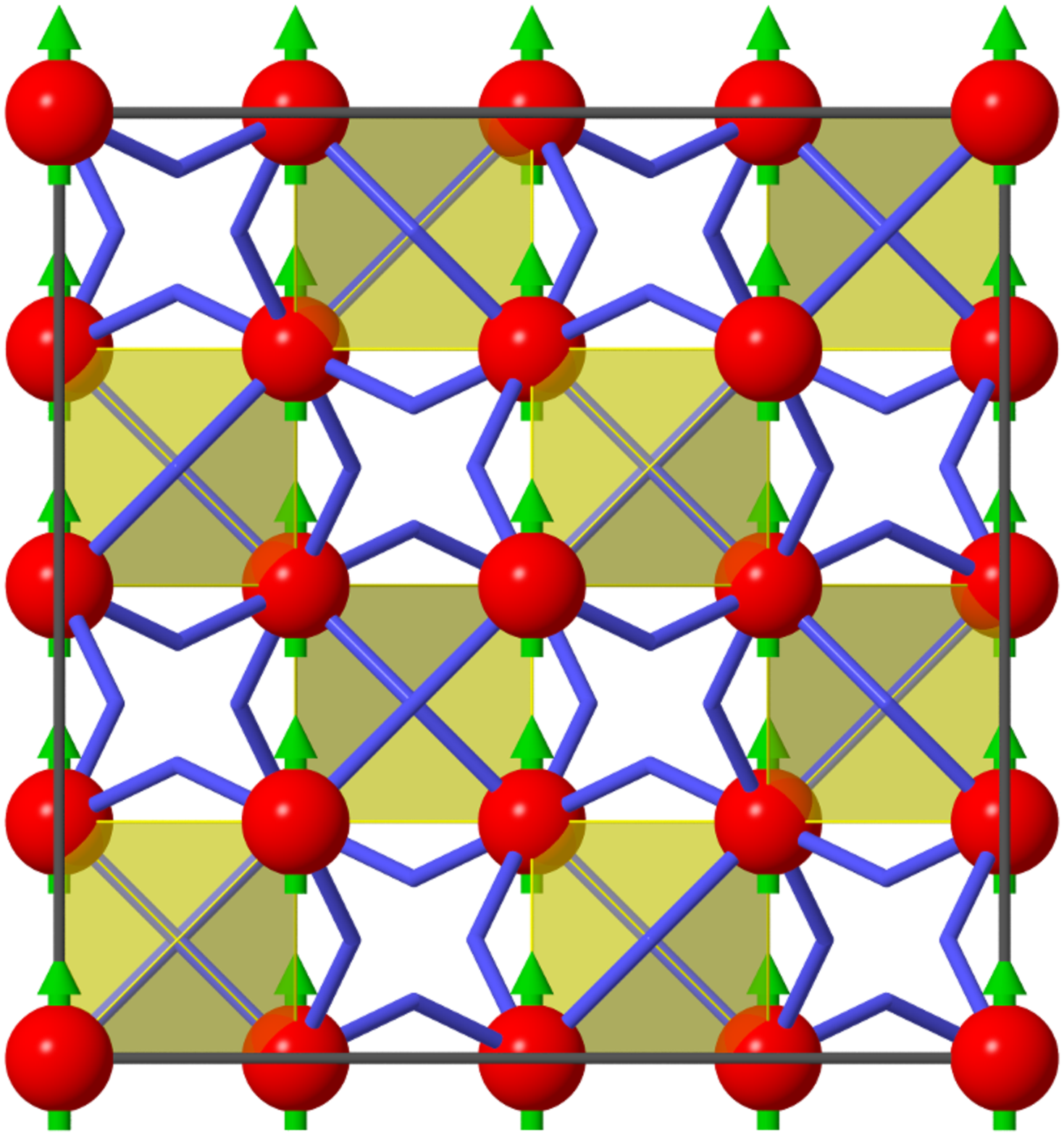} &
\includegraphics[width=0.5\columnwidth]{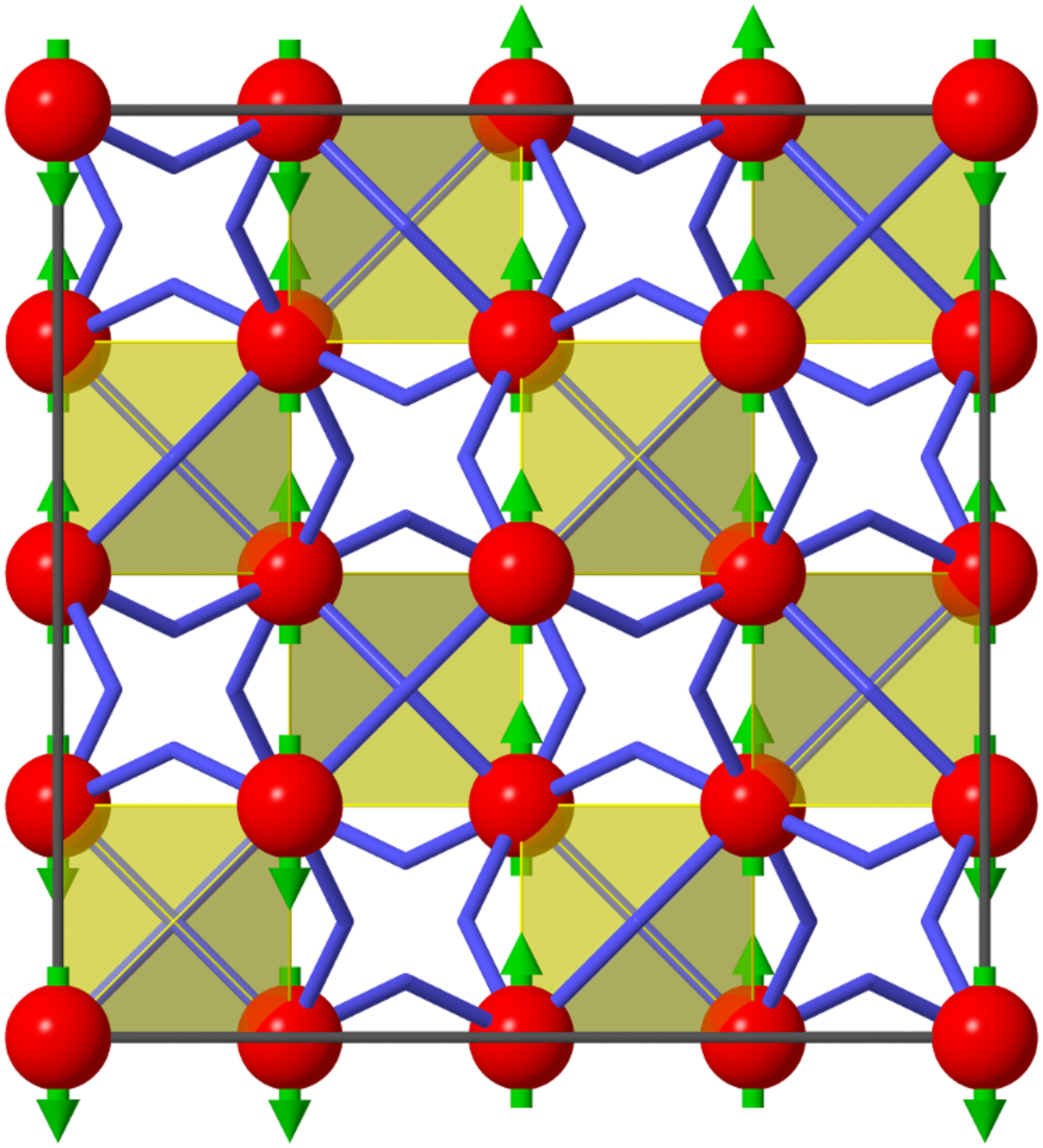} \\
C &
D \\
\includegraphics[width=0.5\columnwidth]{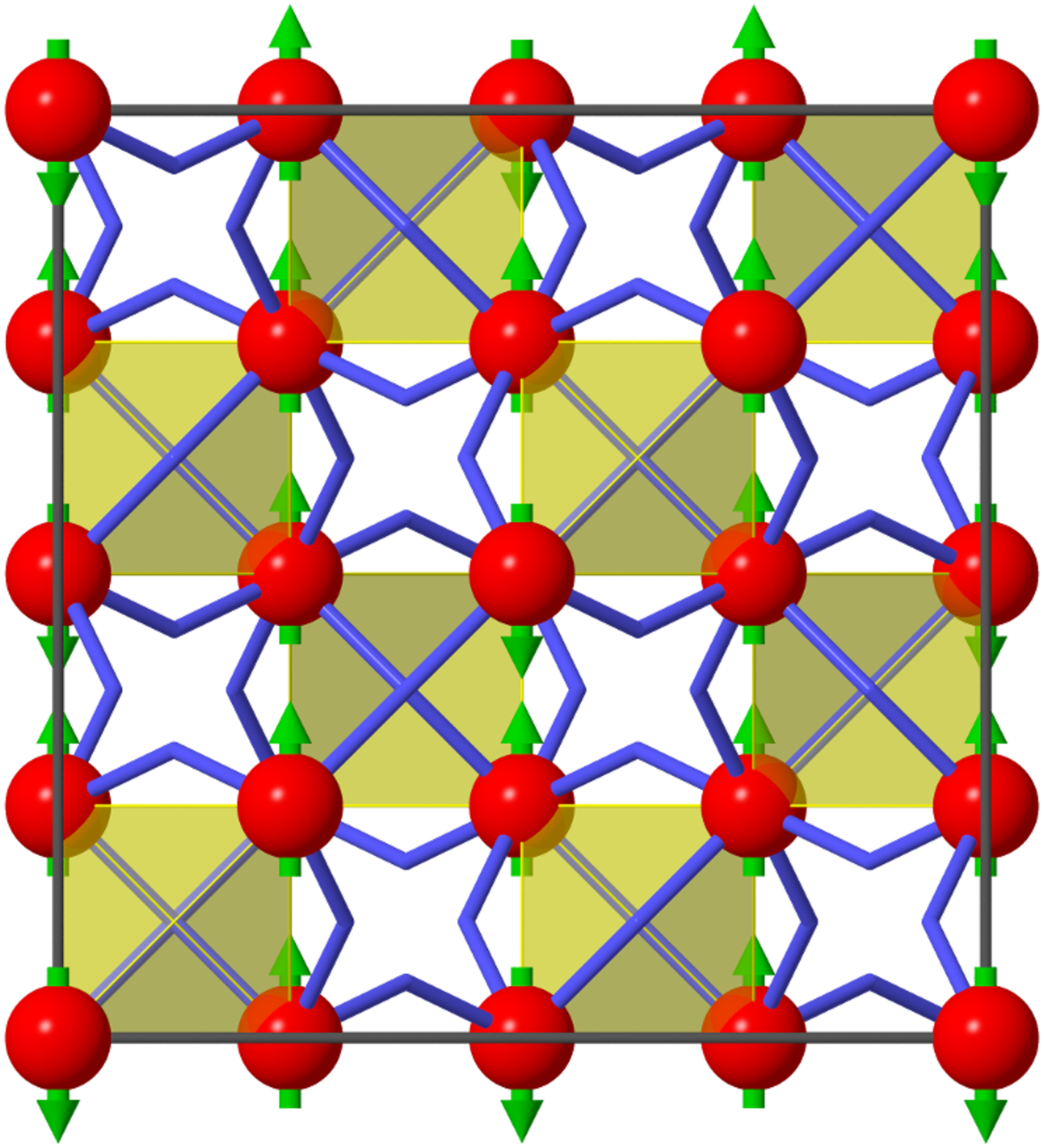} &
\includegraphics[width=0.5\columnwidth]{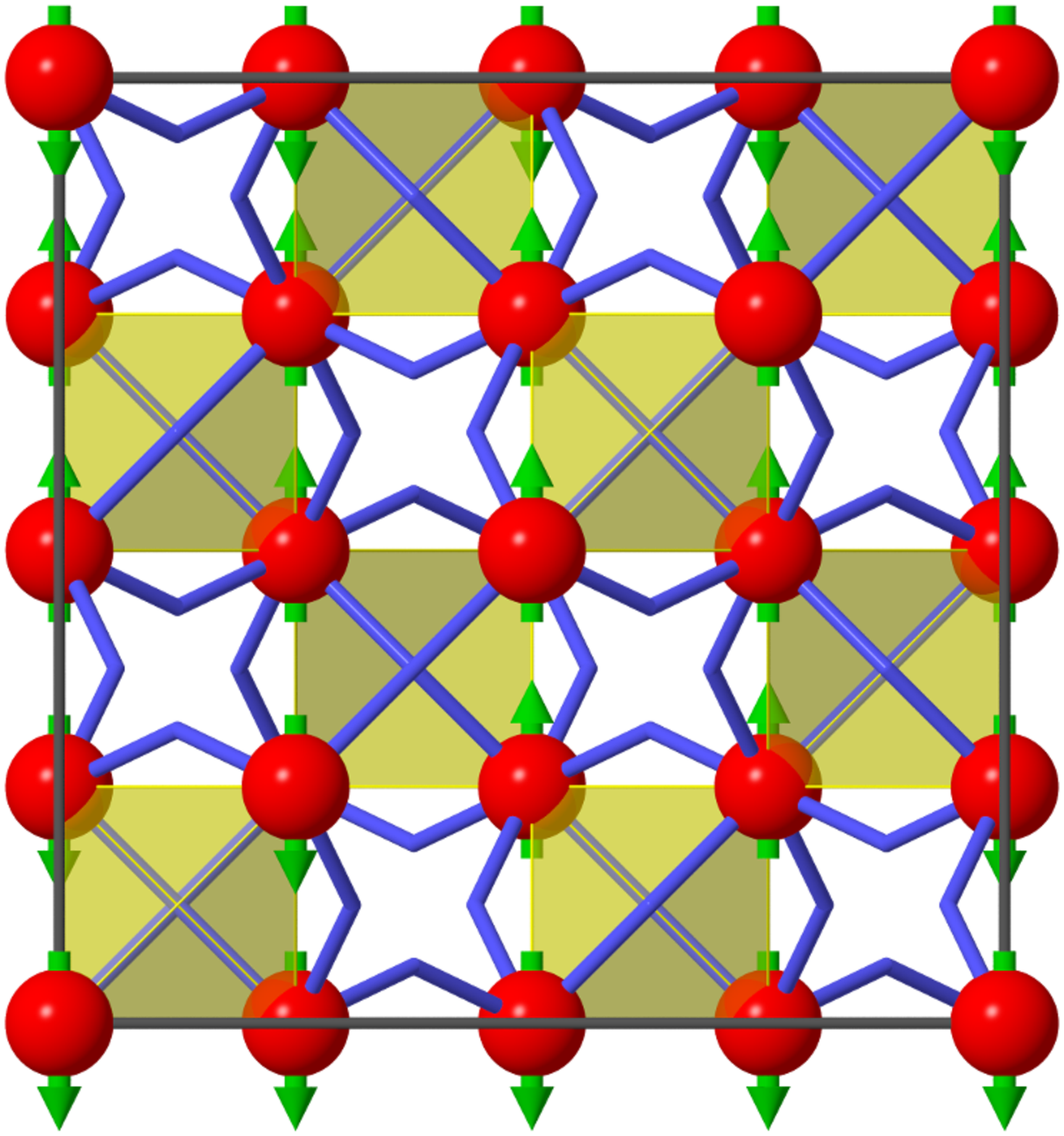} 
\end{tabular}
 \caption{The four collinear magnetic moment configurations (A, B, C and D) to derive the Heisenberg exchange interactions ${J}_{1}$, 
${J}_{2}$,   $J_{3a}$.           The direction of magnetic moments are by the green arrows.}
    \label{fig:spins}
\end{figure}

\subsubsection{bi-quadratic term}
To calculate the nearest-neighbor bi-quadratic coupling $B_1$, we seek   magnetic configurations that are energetically 
degenerate in terms of the Heisenberg interactions, that is  in the absence of spin-orbit correction.
A systematic way of generating  such configurations is as follows.
The direction of the magnetic moments can be characterized by  polar  
  and azimuthal angles  $\theta$  and $\phi$, respectively.  Starting  from an all-in/all-out configuration,  
we choose two Fe ions on a tetrahedron and rotate  the direction of their 
magnetic moment  according  to $\theta'_{1,2}=\theta_{1,2}+ \delta$, $\phi'_{1,2}=\phi_{1,2}-2\delta$. For the remaining two  Fe ions, 
we do the same  but change the sign  of $\delta$, $\theta'_{3,4}=\theta_{3,4} - \delta$, $\phi'_{3,4}=\phi_{3,4} + 2\delta$. 
In this way, the vector sum of the magnetic moments remains equal to zero under this rotation.  
The  contribution from the Heisenberg terms remaining unchanged by this rotation, the only contributions to the variation of the total energy 
come from the other isotropic terms within LDA and LDA+$U$ (the anisotropic terms do not play any role because the spin-orbit coupling is not yet considered at this point).
Ignoring the ring-exchange, $K$, and  fitting the energy variations (obtained by LDA+$U$ with $U_{\rm eff}=2.8$ eV), 
versus $\delta$ (shown in Fig.~\ref{fig:BKC}),   
we find the nearest-neighbor bi-quadratic coupling $B_{1}\approx 1.0$ meV.

\begin{figure}
    \includegraphics[width=\columnwidth]{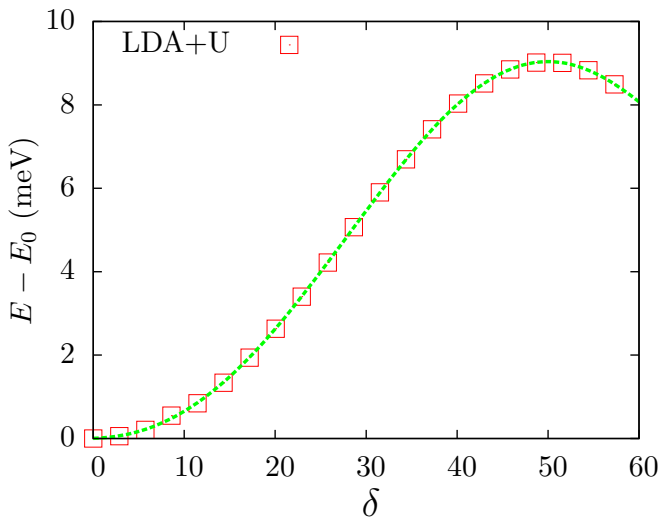}
    \caption{Total energy difference (within LDA+$U$) between all-in/all-out configuration and 
             the configurations with the condition of zero total zero moment,
 ${\bm S}_{\rm tot}=0$, on each tetrahedron. $\delta$ denotes the amount of rotation of spins in each configuration within a tetrahedron  with respect to the all-in/all-out state
($\theta'_{1,2}=\theta_{1,2}+ \delta$, $\phi'_{1,2}=\phi_{1,2}-2\delta$, $\delta$, $\theta'_{3,4}=\theta_{3,4} - \delta$, $\phi'_{3,4}=\phi_{3,4} + 2\delta$).  The dash line shows the fitting to the data using the bi-quadratic term of the Hamiltonian $H_{\rm b.q.}$.}
    \label{fig:BKC}
\end{figure}
\begin{figure}
    \includegraphics[width=\columnwidth]{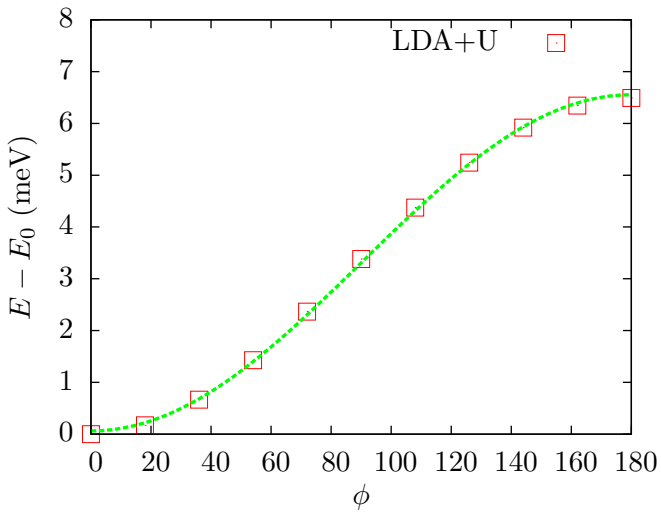}
    \caption{ Total energy difference  (within LDA+$U$) between the all-in/all-out configuration and  the corresponding rotated one  by the azimuthal angle  $\phi$. 
              The dash line shows the fitting to the  data using the DM term in the  Hamiltonian, $H_{\rm DM}$. } 
    \label{fig:DMI} 
\end{figure}
\subsubsection{Single-ion and Dzyaloshinskii-Moriya terms}

Proceeding along, in order to derive the  magnitude  $D$ of the DM interaction,  we require magnetic configurations
for which  the full $O(3)$ isotropic  part of the Hamiltonian,
${\cal H}_{\rm H}+{\cal H}_{\rm b.q.}+{\cal H}_{\rm r}$, remains unchanged.  

In the absence of  spin-orbit coupling (SOC), 
a uniform rotation of  all  magnetic moments 
does not change the total energy due to the  $O(3)$ symmetry of the non-relativistic  part (see main text).  
However, including the spin-orbit correction to the 
{\it ab initio} calculations (LDA+U+SOC), lifts the rotational symmetry.  
We start from an  all-in (all-out) configuration and rotate all the magnetic moments uniformly by an angle $\phi$ around the global $z$-axis.    
Fitting the symmetry breaking relativistic corrections,   
$H_{\rm {DM}}+H_{\rm {s.i.}}$, to  the computational  total DFT energy  (taking $U_{\rm eff}=2.8$ eV) versus 
  $\phi$ (see Fig.~{\ref{fig:DMI}}), enables us to obtain   $D\approx 0.6$ meV and $\Delta\sim 0.0$  for the DM and single-ion couplings, respectively.


\section{Quantum fluctuations and F\lowercase{e} \& F magnetic moments}
\label{sec:QF}

Hybridization between Fe $d$-orbitals   and 
F $p$-orbitals changes the formal ionization state of the Fe and F atoms.
This covalency effect \cite{covalent} results in a weak magnetic moment for 
F ($\mu_{\rm F} \approx 0. 16$  $\mu_{\rm B}$ within the muffin-tin sphere) 
along with a magnetic moment smaller than the full $5$ $\mu_{\rm B}$ 
ionic value for Fe$^{3+}$ ($\mu_{\rm Fe} \approx 4.2$  $\mu_{\rm B}$ in the muffin-tin sphere). 
The experimentally observed Fe long-ranged ordered moment is 3.32(7) $\mu_{\rm B}$ \cite{ferey86-1}, 
presumably reduced from the $4.2$  $\mu_{\rm B}$ LDA+$U$ value by quantum fluctuations \cite{Maestro}.
which can remain sizeable for perturbatively small terms in ${\cal H}$ beyond the nearest-neighbor
exchange $J_1$ \cite{Maestro}. The small $0. 16$  $\mu_{\rm B}$ F 
moments would be further reduced by the quantum fluctuations of the Fe moments to which they are is enslaved to.

To the best of our knowledge, the local magnetic moment ($\sim 0.16$ $\mu_{\rm B}$) 
on the F ion in ${\rm FeF}_3$ has not been measured experimentally. 
In part, this may be because the number of experimental studies carried out on this material has remained 
few until now, and it is a key purpose of our work to motivate new investigations of this compound. 
Perhaps Fluorine ($^{19}$F) nuclear magnetic resonance (NMR)  could shed light on the existence of a local moment
 on F (see Ref.~[\onlinecite{covalent}]).  {As an example of the measurement of the  magnetic moments on anion (ligand) in a transition-metal compound, we note that an oxygen 	
(${\rm O}$) magnetic moment  of  $0.14\mu_{\rm B}$ in $\rm {CuO}$ has been experimentally detected~\cite{CuO}}. 


That being said, one would expect the Fe and F magnetic moments calculated within the LDA+$U$ method 
to be reduced under the quantum spin fluctuations of the Fe moments not included within  LDA+$U$.
For example, the experimentally observed value (via neutron scattering \cite{ferey86-1}) of the Fe magnetic moment is 
$3.32(7)$ $\mu_{\rm B}$ while the calculated LDA+$U$ value is $\sim 4.2$ $\mu_{\rm B}$.  
One could quite naturally ascribe such a  significant  reduction of 20\% 
of the ordered moment to quantum fluctuations of the Fe magnetic moments away from the all-in/all-out ground state as we now explain. 

Because of its broken global discrete symmetry nature, the all-in/all-out (AIAO) ground state stabilized by the Dzyaloshinskii-Moriya (DM) 
interactions would have all magnetic excitations gapped throughout the Brillouin zone and one would therefore naively expect 
the zero point (quantum) fluctuations to be quite small. 
However, because of the two completely flat zero-energy branches of magnons (zero frequency modes) associated with the original 
pure Heisenberg pyrochlore antiferromagnet model \cite{Maestro} describing this material, which then become gapped and weakly dispersive
 once the DM interactions are included, quantum fluctuations need not be  negligible. For example, a calculation of such 
quantum fluctuations was carried out for the pyrochlore Heisenberg antiferromagnet with additional perturbative long-range dipolar interactions 
which also stabilize a broken discrete symmetry ground state \cite{Maestro}.
 A calculation of the zero-point fluctuation-reduction of the moment in FeF$_3$ could be carried out once an accurate
experimental determination of the exchange constants and the DM interaction has been achieved 
(for example from an inelastic neutron scattering measurement of the spin wave dispersion below the N\'eel temperature). 
As for the F ions, the calculated magnetic moment within LDA+$U$ is already quite small
($\sim 0.16$ $\mu_{\rm B}$) and might be difficult to detect experimentally, 
even more so once it is further reduced by quantum fluctuations since the F ions are tied to the Fe ions and their quantum dynamics.

 {In the linear augmented plane wave (LAPW) method, the computed magnetic moments depend on the choice of  the radius of the muffin-tin ($R_{\rm {MT}}$) sphere. Here we investigate  this dependency for the Fe magnetic moments, within LDA+$U$+SOC for the AIAO spin configuration. We show that varying   $R_{\rm {MT}}$ of the Fe ions from 2.2 to 1.8 a.u., results in a slight decrease  of the Fe magnetic moments from $4.24$ to $4.07 \mu_{\rm B}$  (see Table~\ref{tab:FeM}). Therefore, the value of $R_{\rm {MT}}$ does not have a major effect on the Fe magnetic moment in $\rm {FeF}_{3}$. Hence one would still need to account for the quantum fluctuations to fill the gap between the DFT values of $M_{\rm {Fe}}$ and  the experimental one.  
It has to be noticed that to maintain a good  accuracy in the calculations, we have to enlarge the $K_{\rm {max}}$ when decreasing  $R_{\rm {MT}}$, which we have done.}

\begin{table}[h]
\caption{Dependence of the Fe magnetic moment on muffin-tin radius ($R_{\rm {MT}}$)  within LDA+U+SOC.}
\label{tab:FeM}
\begin{tabular}{|c|c|}
\hline
        $R_{\rm {MT}}$ (a.u.)  &    $M_{\rm {Fe}}$ $(\mu_{\rm B})$  \\ \hline
       1.8         &  4.07         \\ \hline
       1.9         &  4.12         \\ \hline
       2.0         &  4.17         \\ \hline
       2.1         &  4.21         \\ \hline
       2.2         &  4.24         \\
\hline
\end{tabular}
\end{table}


{We now compare the results obtained using the generalized gradient approximation (GGA) with those obtained by LDA.
 For the  AIAO configuration, the LDA magnetic moment of Fe is found to be $\sim 4.03$ $\mu_{\rm B}$. Using GGA for the exchange-correlation functional, the magnetic moment of
Fe is determined to be $\sim 4.13$ $\mu_{\rm B}$. For this calculation we chose (Perdew-Burke-Ernzerhof) PBE functional~\cite{PBE}. Hence the usage of the GGA does not result in a significantly different  Fe  magnetic moment compared to LDA.}


\section{Dependence of the results on $U$, $J_{\rm H}$ and $U_{\rm eff}$ }
\label{sec:DFT_dep}

To investigate the robustness of the results 
presented in the main text upon changing  the values of the on-site Coulomb interaction $U$, the Hund's exchange  $J_{\rm H}$ and the
effective Coulomb interaction $U_{\rm eff} \equiv U-J_{\rm H}$, we performed further  DFT calculations  by choosing  different values of these  parameters. 
{First we should mention that in the DFT scheme we used, $U$ and $J_{\rm H}$, enters separately in the energy functional of LDA+U.}
Hence, we start with the fixed value  $U_{\rm eff}=2.8\, {\rm eV}$ and change the values of $J_{\rm H}$ and $U$ accordingly. 
The resulting couplings of the spin Hamiltonian as well as the Fe magnetic moment within the muffin-tin sphere, for $J_H=0.5, 0.75, 1.0$ and $1.25$ eV,
are given in Table.~\ref{tab:JH}. It is evident from this table that the results are robust against the variation of $J_{\rm H}$,
 provided that the effective Coulomb interaction $U_{\rm eff}$ remains constant.  
We should mention, however, that such independence of the microscopic quantities with $U_{\rm eff}=U-J_{\rm H}$ kept constant as $U$ and
$J_{\rm H}$ are independently varied is not found in all systems.
For example, consider Fig. 2 in Ref.~[\onlinecite{Spaldin}] in which the magnetocrystalline anisotropy energy (MCAE) of FeF$_2$ is computed
for fixed $U_{\rm eff}$ while two pairs of $U$ and $J_{\rm H}$ values are used;
$U=6.0$ eV and $J_{\rm H}=1.2$ eV and $U=5.0$ eV and $J_{\rm H}=0.2$ eV, which both give $U_{\rm eff}=4.8$ eV.
The pair ($U=6$, $J_{\rm H}=1.2$) gives the correct experimentally measured MCAE while the
 pair    ($U=5$ and $J_{\rm H}=0.2$) gives a completely different and therefore incorrect MCAE in comparison with experiment.

Next, we proceed to check the dependence of the results upon a
 variation of $U_{\rm eff}$. For this purpose we used 
$U_{\rm eff}=4.0$ eV (i.e. $U=5$ and $J_{\rm H}=1.0$) and 6.0 eV (i.e.  $U=7$ and $J_{\rm H}=1.0$) in addition to the value 2.8 eV
obtained by the linear response approach (see text in main paper).
The computed parameters  are listed in Table.~\ref{tab:Ueff}.
These show that the results are quite sensitive to $U_{\rm eff}$, in a sense that $J_1$,
$D$ and $B$ decrease by increasing the value of effective on-site Coulomb interaction $U_{\rm eff}$. 
For completeness the  Monte Carlo results of  the Hamiltonians  obtained for  $U_{\rm eff}=4.0$ and $6.0\,{\rm eV}$ are  presented in Figs.~\ref{fig:U4} and \ref{fig:U6}, respectively. 
One sees from  Figs.~\ref{fig:U4}a and \ref{fig:U6}a that, as the transition temperature is decreased by increasing $U_{\rm eff}$, 
the discontinuous nature of the transition becomes more pronounced. 
This is consistent with the statement made in the main text that has $D/J_1$ is decreased, and the system approaches the isotropic Heisenberg antiferromagnet limit, the transitions become progressively more strongly first order. 

The probability distributions for the AIAO order parameter $m_n$ 
(Figs.~\ref{fig:U4}b and \ref{fig:U6}b) and the distribution function for the 
four spin correlation $R$, $P(R)$ in (Figs.~\ref{fig:U4}c and \ref{fig:U6}c),
confirm that the general picture of coexisting AIAO and 
 co-planar states in the vicinity of the transition point, discussed in the 
main body of the letter, is still operating and, therefore,  does not hinge on a precise choice of $U_{\rm eff}$.

\begin{table*}[t]
\caption{Parameters obtained from ab intio calculations (LDA+$U$) with different values of $J_H$, and constant $U_{\rm eff}=2.8$ eV.
$J_{1}$, $J_{2}$ and $J_{{\rm 3a}}$ are the first,  second and third  neighbor exchange interactions, respectively. $B_1$ and $D$ denote
the bi-quadratic and   Dzyaloshinskii-Moriya couplings, respectively. $M_{\rm Fe}$ is the magnetic moment of the Fe ion in the muffin-tin sphere.
}
\label{tab:JH}
\begin{ruledtabular}
\begin{tabular}{c|ccccccc}
      $J_{\rm H}$ $({\rm eV})$   &  $U$ $({\rm eV})$     &   $J_1$ $({\rm meV})$ &  $J_2$ $({\rm meV})$ &   $J_{3a}$ $({\rm meV})$    &  $B_1$ $({\rm meV})$   &  $D$ $({\rm meV})$  &  $M_{\rm Fe}$ (${\mu_{B}}$) \\ \hline
       0.5         &       3.30    &  32.484       &   0.590      &       0.541         &  0.967       &     0.567   &     4.244     \\ \hline
       0.75        &       3.55    &  32.607       &   0.591      &       0.541         &  0.961       &     0.570   &     4.244      \\ \hline
       1.00        &       3.80    &  32.731       &   0.592      &       0.540         &  0.954       &     0.573   &     4.243      \\ \hline
       1.25        &       4.05    &  32.861       &   0.592      &       0.541         &  0.948       &     0.576   &     4.243      \\
\end{tabular}
\end{ruledtabular}
\end{table*}

\begin{table*}[t]
\caption{Parameters of the spin Hamiltonian obtained  by ab intio calculations (LDA+$U$) using $U_{\rm eff}=2.8, 4.0, 6.0$ eV.
The last column shows the transition 
temperatures obtained from Monte Carlo simulations for a system size with $N=4\times 6^3$
}
\label{tab:Ueff}
\begin{ruledtabular}
\begin{tabular}{c|cccccc}
$U_{\rm eff} ({\rm eV})$        &  $J_1\, ({\rm meV})$   &  $D/J_1$ &  $J_2/J_1$ &  $J_{3a}/J_1$ & $B_1/J_1$  &  $T_c$  (K)        \\ \hline
       2.8                            &  32.7                  &  0.018      &   0.018         &  0.015              &  0.030       &    $\sim 22$           \\ \hline    
      4.0                            &  27.4                   &   0.011     &   0.018         &  0.010              &  0.018        &  $\sim 11.5$           \\ \hline    
       6.0                           &  20.9                  &    0.005    &    0.014         &  0.010             &  0.010         &   $\sim 4.2$              \\  
\end{tabular}
\end{ruledtabular}
\end{table*}


\begin{figure*}[t]
    \includegraphics[width=\textwidth]{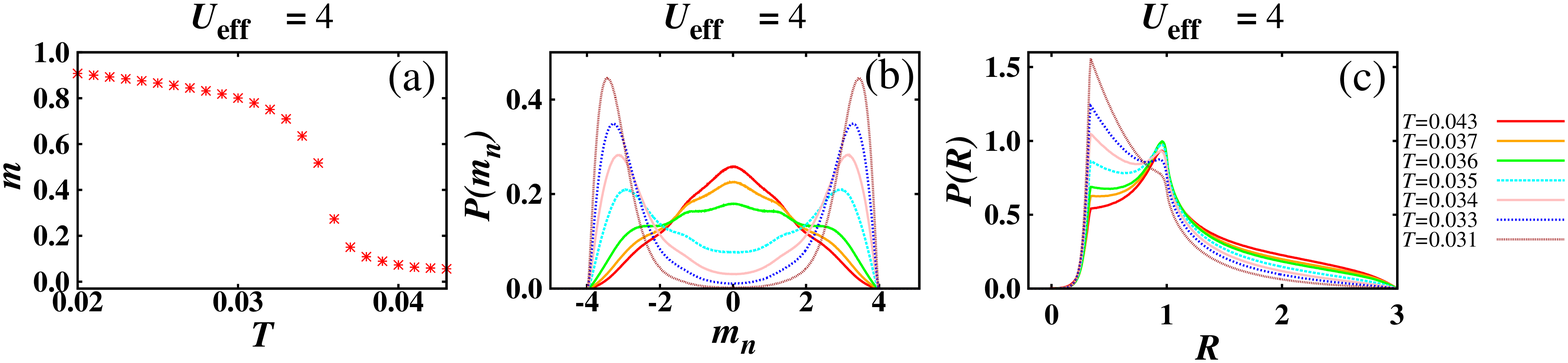}
    \caption{ (a) Variation of the AIAO order parameter, $m$, versus 
temperature $T$ (in units of $J_1$).
(b) Probability distribution functions for
the AIAO order parameter per tetrahedron, $m_n$.
(c) Four spin correlation per tetrahedron, $R$, as defined by Eq.~\ref{R} in below,    
for a lattice of linear size $L=6$  and $U_{\rm eff}=4.0$ eV.    } 
    \label{fig:U4} 
\end{figure*}
\begin{figure*}[t]
    \includegraphics[width=\textwidth]{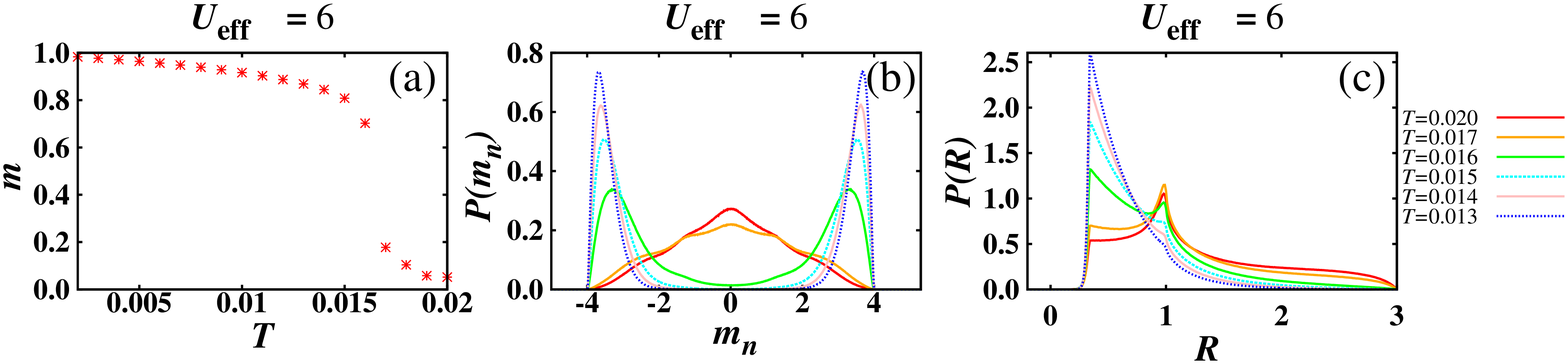}
 \caption{ (a) Variation of the AIAO order parameter, $m$, versus 
temperature $T$ (in units of $J_1$).
(b) Probability distribution functions for
the AIAO order parameter per tetrahedron, $m_n$.
(c) Four spin correlation per tetrahedron, $R$, as defined by Eq.~\ref{R}   
for a lattice of linear size $L=6$  and $U_{\rm eff}=6.0$ eV.    }     
    \label{fig:U6} 
\end{figure*}


\section{energy landscape for a single tetrahedron}

\label{sec:biquad}

In this section, we compute the classical ground state energy of the AIAO and coplanar states.
We consider four classical spins residing on the corners of a single tetrahedron, interacting via  a  
Hamiltonian ${\cal H}$ that includes the  antiferromagnetic Heisenberg, bi-quadratic and DM interactions.

\begin{equation}
{\cal H}={\cal H}_{\rm H} + {\cal H}_{\rm b.q.} + {\cal H}_{\rm DM}.
\end{equation} 
The first term has a highly degenerate ground state manifold characterized by  the two angles $\theta$ and $\phi$ as shown in 
Fig.~\ref{gs-tetrahedron}.  
Choosing the $z$-axis along ${\bm S}_{3}+{\bm S}_{4}$ (the dashed line in Fig.~\ref{gs-tetrahedron}), 
we can parametrize  the spins within a tetrahedron by $\theta$ and $\phi$

\begin{eqnarray}
\label{spin-parameter}
{\bm S}_1&=&(-\cos \frac{\theta}{2} , 0 ,\sin \frac{\theta}{2})\nonumber \\
{\bm S}_2&=&(\cos \frac{\theta}{2} , 0 ,\sin \frac{\theta}{2})\nonumber\\
{\bm S}_3&=&(\cos \frac{\theta}{2} \cos \phi , \cos \frac{\theta}{2} \sin \phi , -\sin \frac{\theta}{2}) \nonumber\\
{\bm S}_4&=&(-\cos \frac{\theta}{2} \cos \phi , -\cos \frac{\theta}{2} \sin \phi , -\sin \frac{\theta}{2}). 
\end{eqnarray}
The above relations enable us to write the bi-quadratic term as follows:

\begin{equation}
{\cal H}_{\rm b.q.}=B_{1}\cdot Q,
\end{equation}
in which 

\begin{eqnarray}
Q &\equiv& \sum_{<i,j>} ({\bm S}_i \cdot {\bm S}_j)^2\\\nonumber
&=&1-2\sin ^2 \phi \cos \theta +(3+\cos^2 \phi) \cos ^ 2\theta +\cos^2\phi.
\end{eqnarray}
The quantity $Q$ as a function of $\theta$ and $\phi$,  depicted in  Fig.~\ref{bq-landscape},
shows a minimum at  $\phi=\frac{\pi}{2}$ , $\theta=\cos^{-1}(1/3)$ with value $Q=2/3$. 
This means that the ground state configuration corresponds to a non-coplanar state with  an angle 
of $109.47^{\circ}$ between each pair of spins. 
In this configuration, the plane of each pair is perpendicular to the plane of the other two spins. 
The $O_3$ symmetry of the bi-quadratic interaction offers the freedom to  rotate this configuration 
by any arbitrary angle. However, it can be easily seen that the direct DM interaction selects 
the orientation in which the  spins are aligned along the vectors connecting the 
corners to the center of the tetrahedron, the so called all-in/all-out (AIAO) configuration \cite{elhajal}.  

Crucially, Fig.~\ref{bq-landscape}  shows {three saddle points} for $Q$ at
 $\{\phi=0 , \theta=\frac{\pi}{2}\}$, $\{\phi=\pi , \theta=\frac{\pi}{2}\}$ and $\{\phi=\pi/2 , \theta=0\}$. {These saddle points  correspond to  the coplanar states}} discussed in the main body of the paper.
 There are  three independent choices for constructing such a state. Depending on which two spins are
 considered to be collinear, the DM interaction restricts the spins to be in one of the $xy$, $xz$ or $yz$ planes.  To show this, first assume   
${\bm  S}_{1}=-{\bm  S}_{2}$ ,  ${\bm  S}_{3}=-{\bm  S}_{4}$ and ${\bm S}_{1}\perp {\bm S}_{3}$. 
Then,  using Eqs.~(\ref{DM-vectors}) below, we find for the DM term
\begin{eqnarray}
&&\sum_{ij} {\bf D}_{ij} \cdot ({\bm S}_{i}\times {\bm S}_{j})\\\nonumber
&&=({\bf D}_{13}+{\bf D}_{24}-{\bf D}_{14}-{\bf D}_{23})\cdot ({\bm S}_{1}\times {\bm S}_{3})\\\nonumber
&&=-2{\sqrt{2}} {\hat {\bf e}_{x}}\cdot ({\bm S}_{1}\times {\bm S}_{3}) .
\end{eqnarray} 
From this, the  minimum energy condition requires that ${\bm S}_{1}$ and $ {\bm S}_{3}$ lay in the $yz$-plane in such 
a way that their cross product gives ${\bf e}_{x}$.  If we choose 
${\bm  S}_{1}=-{\bm  S}_{3}$ ,  ${\bm  S}_{2}=-{\bm S}_{4}$ and ${\bm S}_{1}\perp {\bm S}_{2}$, we find
\begin{eqnarray}
&&\sum_{ij} {\bf D}_{ij} \cdot ({\bm S}_{i}\times {\bm S}_{j})\\\nonumber
&&=({\bf D}_{12}-{\bf D}_{14}+{\bf D}_{23}+{\bf D}_{34}) \cdot ({\bm S}_{1}\times {\bm S}_{2})\\\nonumber
&&=2{\sqrt{2}} {\hat {\bf e}_{y}}\cdot ({\bm S}_{1}\times {\bm S}_{2}) .
\end{eqnarray} 
From this, we find  that ${\bm S}_{1}$ and $ {\bm S}_{2}$ lay in the $xz$-plane in such a way that their cross product gives $-{\bf e}_{y}$. 
Finally, taking   ${\bm  S}_{1}=-{\bm  S}_{4}$,  ${\bm  S}_{2}=-{\bm  S}_{3}$ and ${\bm S}_{1}\perp {\bm S}_{2}$, we get
\begin{eqnarray}
&&\sum_{ij} {\bf D}_{ij} \cdot ({\bm S}_{i}\times {\bm S}_{j})\\\nonumber
&&({\bf D}_{12}-{\bf D}_{13}+{\bf D}_{24}-{\bf D}_{34})\cdot ({\bm S}_{1}\times {\bm S}_{2})\\\nonumber
&&2{\sqrt{2}} {\hat {\bf e}_{z}}\cdot ({\bm S}_{1}\times {\bm S}_{2}),
\end{eqnarray} 
which implies  that ${\bm S}_{1}$ and $ {\bm S}_{2}$ lay in the $xy$-plane in such a way that their cross product gives $-{\bf e}_{z}$.    

The above arguments lead us to the following expressions for the energy per spin of the coplanar and all-in/all-out (AIAO) states,
$\epsilon_{\rm coplanar}$ and $\epsilon_{\rm AIAO}$, respectively, for each tetrahedron 
\begin{eqnarray}
\epsilon_{\rm coplanar}	&=&	-J_{1}+B_{1}-D\sqrt{2}		\\
\epsilon_{\rm AIAO} 		&=&	-J_{1}+B_{1}/3-2D\sqrt{2},
\end{eqnarray}
hence we have 
\begin{equation}
\label{cadiff}
\epsilon_{\rm coplanar}-\epsilon_{\rm AIAO}=(\frac{2B_1}{3}+D\sqrt{2})>0,
\end{equation}
that is the ground state is AIAO for all $B_1>0$ and $D>0$.

\begin{figure}[t]
\centering 
\includegraphics[width=0.8\columnwidth]{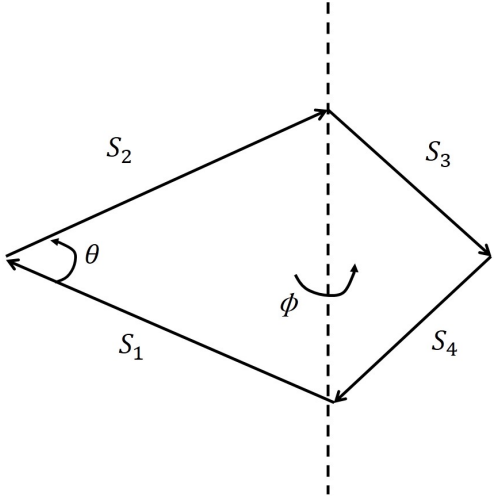}
\caption{A ground state configuration of four classical spins on a tetrahedron and coupled by antiferromagnetic Heisenberg interaction.  }
\label{gs-tetrahedron}
\end{figure}

\begin{figure}[t]
\centering 
\includegraphics[width=\columnwidth]{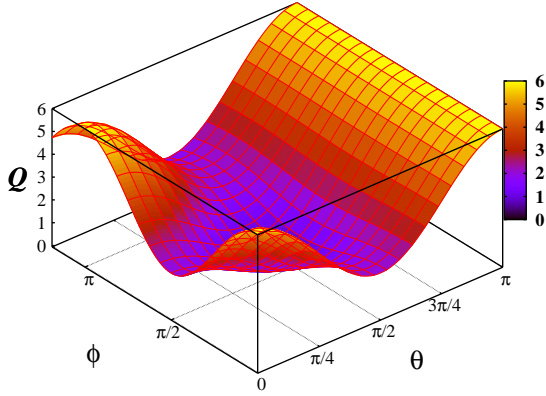}
\caption{Energy landscape function $Q$ of the  bi-quadratic part of ${\cal H}$ alone, 
$H_{\rm b.q.} = B_1 Q(\theta,\phi)$,
in terms of  $\theta$ and $ \phi$.
}
\label{bq-landscape}
\end{figure}


\section{Dzyaloshinskii-Moriya (DM) vectors}
\label{sec:DM} 

\begin{figure}[t]
\includegraphics[width=\columnwidth]{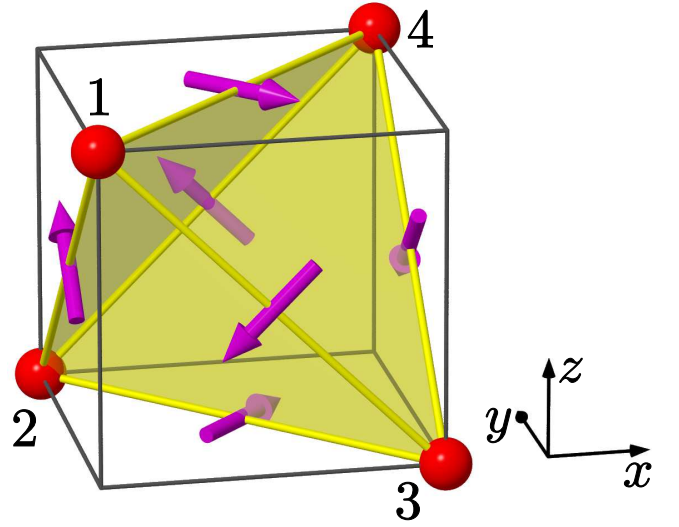}
\caption{Orientation of the DM vectors for a single tetrahedron.}
\label{fig:dm}
\end{figure}

The following minimal spin Hamiltonian, ${\cal H}_{\rm min}$, is considered in the main part of the paper:

\begin{eqnarray}\label{H}
{\cal H}_{\rm min}&=&{J_1}\sum_{<i,j>}\sum_{a,b}{\bm S}_{i}^{a}\cdot{\bm S}_{j}^{b}\\\nonumber
&+&{B_1}\sum_{<i,j>}\sum_{a,b}({\bm S}_{i}^{a}\cdot{\bm S}_{j}^{b})^2\\\nonumber
&+&{D}\sum_{<i,j>}{\bf D}^{ij}_{ab}\cdot({\bm S}_{i}^{a}\times{\bm S}_{j}^{b}),
\end{eqnarray}
in which  the moment ${\bm S}_i$ is a classical unit vector,  $J_{1} > 0$ is the nearest-neighbor antiferromagnetic exchange interaction, 
$B_{1}>0 $  is the nearest-neighbor bi-quadratic interaction while the last term is the anisotropic DM interaction. 
$i,j=1\cdot\cdot N$ and
$a,b=1,2,3,4$ denote the Bravais lattice and sub-lattice indices,
respectively and $\langle i,j \rangle$ means the nearest-neighbor lattice sites. Considering a single tetrahedron, the plane which
contains two neighboring lattice points  and the  middle-point of the opposite
bond in the tetrahedron is a mirror plane. Applying Moriya's rules~\cite{moriya}
implies that the {\bf D} vector can only be perpendicular to
this mirror plane or, equivalently, parallel to the opposite bond.
Therefore,  ${\bf D}^{ij}_{ab}$'s represent the vectors along the six directions given by:

\begin{eqnarray}\label{DM-vectors}
{\bf D}_{12}&=&{D\over \sqrt{2}}(0,1,1)\nonumber\\
{\bf D}_{13}&=&{D\over \sqrt{2}}(-1,0,-1)\nonumber\\
{\bf D}_{14}&=&{D\over \sqrt{2}}(1,-1,0)\nonumber\\
{\bf D}_{23}&=&{D\over \sqrt{2}}(1,1,0)\nonumber\\
{\bf D}_{24}&=&{D\over \sqrt{2}}(-1,0,1)\nonumber\\
{\bf D}_{34}&=&{D\over \sqrt{2}}(0,1,-1)
\end{eqnarray}
The orientation of the DM vectors is illustrated in Fig. \ref{fig:dm}.

 There are therefore two possible values for the DM interactions between two nearest-neighbor sites and which correspond to
the two directions for the {\bf D} vector (and keeping the same
order for the cross product 
${\bm S}_{i}^{a} \times {\bm S}_{j}^{b}$),
the ``direct'' DM interaction for $D>0$ and the ``indirect'' one for $D<0$ (Ref.~[\onlinecite{elhajal}]).
For FeF$_3 $, we find from {\it ab initio} DFT calculations that  $D>0$ and that the 
DM interaction is therefore of the direct type.




\section{Energy Binder ratio}
\label{sec:BinderUE}

\begin{figure}[t]
\includegraphics[width=\columnwidth]{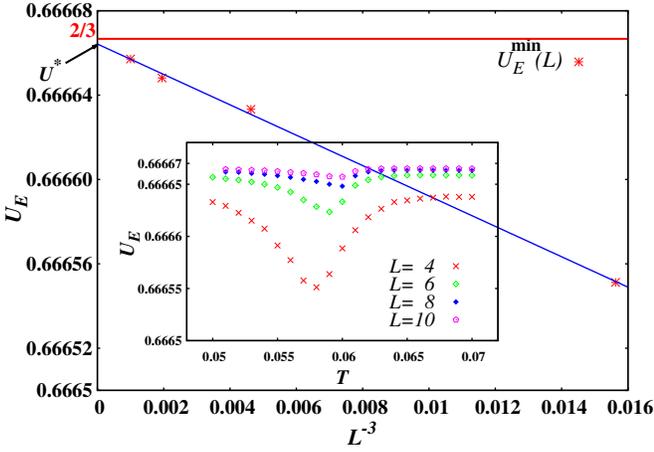}
\caption{ Main panel: scaling of the minimum of the  Binder's fourth cumulant of energy, $U_E^{\rm min}(L)$ versus $1/L^3$.
 Inset: variation  of  $U_E$  versus  temperature, $T$ 
(in units of $J_1$), for lattices of linear size $L=4,6,8,10 $. }
\label{E-binder}
\end{figure}

The energy Binder ratio, $U_E(T)$, was employed in the Monte Carlo simulations to assess the order of the transition to the AIAO long-range ordered state. $U_E(T)$ is defined as
\begin{eqnarray}\label{Binder}
U_{E}(T) \equiv  1-\frac{1}{3}\frac{\langle E^4 \rangle}{ {\langle E^2 \rangle}^2}.
\end{eqnarray}
$U_E$ tends asymptotically to $2/3$ in both 
the ordered and paramagnetic phase while reaching a minimum, 
$U_E^{\rm min}(L)$, in the  region near the transition point. 
For a first order transition, the finite-size scaling of 
$U_E^{\rm min}(L)$ is given by \cite{binder3}
\begin{equation}
\label{U-L}
U_{E}^{\rm min}(L)=U^*+AL^{-d}+{\cal O}(L^{-2d}) ,
\end{equation}  
with  $U^* < 2/3 $ and where $d=3$ is the space dimension and
$A$ is a constant.
Figure \ref{E-binder}, which illustrates a precise linear fit of  $U^{\rm min}_{E}(L)$ versus  $L^{-3}$  
with $U^*=0.666664(1)$, hence very close to $2/3$, 
suggests that the transition might actually be very weakly first order \cite{binder3}.

\section{Verification of  the Coplanar correlations above ${\bm T}_c$}
\label{coplanar}

In this section, we present further evidence for the existence short range coplanar correlations above $T_c$. 
In the main text we introduced  a quantity $R$, defined as 

\begin{equation}
\label{R}
R \equiv ({\bm S}_{1}\cdot{\bm S}_{2})({\bm S}_{3}\cdot{\bm S}_{4})
+({\bm S}_{1}\cdot{\bm S}_{3})({\bm S}_{2}\cdot{\bm S}_{4})
+ ({\bm S}_{1}\cdot{\bm S}_{4})({\bm S}_{2}\cdot{\bm S}_{3}),
\end{equation}  
whose probability distribution function (PDF) has a peak at $R \approx 1$ above $T_{c}$. To assess whether this peak  solely corresponds to coplanar states, we introduce another  quantity, ${\tilde R}$, which is independent of $R$, and  is defined within each tetrahedron  as follows 
\begin{equation}
\label{RP}
{ {\tilde R}}\equiv |({\bm S}_{1}\cdot{\bm S}_{2})({\bm S}_{3}\cdot{\bm S}_{4})
- ({\bm S}_{1}\cdot{\bm S}_{3})({\bm S}_{2}\cdot{\bm S}_{4})
+ ({\bm S}_{1}\cdot{\bm S}_{4})({\bm S}_{2}\cdot{\bm S}_{3})|.
\end{equation}  
For $T \ll J_{1}$, it is highly probable that the spins within a tetrahedron are in a configuration for which $\sum_{i=1}^{4} {\bm S}_{i}\approx 0$.  We can then use the spin parametrisation in terms of the two internal (angular) degrees of freedom given by Eq.~(\ref{spin-parameter}). Substituting for the spins from Eq.~(\ref{spin-parameter}) in Eqs.~(\ref{R}) and (\ref{RP}), we obtain the following two equations  

\begin{eqnarray}
\label{R'}
&&R=\frac{1}{2}\Big[1-2\sin^2\phi\cos \theta+(3+\cos^2\phi)\cos^2\theta+\cos^2\phi\Big],\nonumber\\
&&{\tilde R}=|1-\sin^{2}\theta(1+\cos\phi)|.
\end{eqnarray} 

The PDF  of ${\tilde R}$ ($P({\tilde R})$) for $L=10$, shown in Fig. 3c in the main text, 
displays  a peak near ${{\tilde R}}=1$ above $T_c$. For an AIAO state, the values of the two above quantities are  $R=1/3$ and ${\tilde R}=1/9$. 
 {It is easy to show that the two equations $R=1/3$ and ${\tilde R}=1$ have \emph{no common} solutions in terms of $\theta$ and $\phi$ and, therefore the AIAO state has no contribution causing a  peak at ${\tilde R}=1$.   This allows us to conclude that the states giving rise to  ${ R}=1$ are the same as those contributing in the peak corresponding to ${\tilde R}=1$. This  then leaves us with the two  equations $(R=1; {\tilde R}=1)$,  which have the three  common solutions $\{(\phi=\pi) ; (\theta=\frac{\pi}{2})\}$, $\{(\phi=\pi/2) ; (\theta=0)\}$ and $\{(\phi=0) ; (\theta=\frac{\pi}{2})\}$.  This argument evidently proves that the states coexisting with the AIAO states above $T_c$ {\it are} the coplanar $xy$, $xz$ and $yz$ 
spin configurations discussed in Sec.~\ref{sec:biquad}}.   


\section{Neutron scattering Structure function}\label{SF}
\label{Sq}

In this section, we calculate the neutron scattering structure function  defined as
\begin{equation}
S({\bf q})=\sum_{i,{\mu};j,{\nu}} \langle {\bm S}^\perp _{i,{\mu}} \cdot{\bm S}^\perp _{j,{\nu}} \rangle \exp[i{\bf q}\cdot({\bf R}_{i,{\mu}}-{\bf R}_{j,{\nu}})],
\label{s-function}
\end{equation}
where ${\bf q}$ is the wave-vector transfer of the scattered neutron, ${\bm S}^\perp _{i,{\mu}} ={\bm S}_{i,{\mu}}-\frac{({\bm S}_{i,{\mu}}\cdot{\bf q})}{({\bf q}\cdot{\bf q})}{\bf q}$ is the  component of the spin ${\bm S}_{i,{\mu}}$, perpendicular to ${\bf q}$ and   $\langle \cdots \rangle $ denotes the thermal averaging. ${\bf R}_{i,{\mu}}={\bf T}_{i}+{\bf d}_{\mu }$ is  the position of  each of the four  spins $\mu=1,2,3,4$ in the tetrahedron  unit cell $i=1\cdots N_{\rm cell}$ , where ${\bf T}_i$'s are the set of  primitive translation vectors of the fcc Bravais lattice. ${\bf d}_{\mu}$ gives the position of the spins within a tetrahedron and $N_{\rm cell}$ denotes the total number of unit cells which is equal to $N/4$, with $N$ being the total number of spins. With the geometry shown in Fig.~\ref{fig:dm} above, the vectors ${\rm d}_{\mu}$ are given by
\begin{eqnarray}
{\rm d}_{1}&=&1/4(-1,-1,0)\nonumber\\
{\rm d}_{2}&=&1/4(-1,0,-1)\nonumber\\
{\rm d}_{3}&=&1/4(0,-1,-1)\nonumber\\
{\rm d}_{4}&=&{\bf 0}.
\end{eqnarray}

Eq.~\ref{s-function} can be rewritten as  $S({\bf q})=\langle F({\bf q})F^{*}({\bf q})\rangle$, in which $F({\bf q})$ is given by

\begin{equation}
F({\bf q})=\sum_{i=1}^{N_{\rm cell}}\sum_{\mu=1}^{4}\left({\bm S}_{i,{\mu}}-\frac{({\bm S}_{i,{\mu}}\cdot{\bf q})}{({\bf q}\cdot{\bf q})}{\bf q}\right)\exp\Big[i{\bf q}\cdot({\bf T}_{i}+{\bf d}_{\mu})\Big].
\end{equation}

In the case of long range AIAO ordering, in which all the tetrahedra have the same spin configuration, i.e

\begin{eqnarray}
{\bm S}_{1}&=&{1\over \sqrt{3}}(1,1,-1)\nonumber\\
{\bm S}_{2}&=&{1\over \sqrt{3}}(1,-1,1)\nonumber\\
{\bm S}_{3}&=&{1\over \sqrt{3}}(-1,1,1)\nonumber\\
{\bm S}_{4}&=&{1\over \sqrt{3}}(-1,-1,-1),
\end{eqnarray} 
we have $F({\bf q})=N_{\rm cell}f({\bf q})\delta_{{\bf q},{\bf G}}$, where ${\bf G}=2\pi(h,k,l)$ (with $h, k, l$ being integer) denotes the fcc reciprocal  lattice vectors and  $f({\bf q})$ is the unit cell magnetic form factor defined as 

\begin{equation}
\label{fq}
f({\bf q})=\sum_{\mu=1}^{4}\left({\bm S}_{{\mu}}-\frac{({\bm S}_{{\mu}}\cdot{\bf q})}{({\bf q}\cdot{\bf q})}{\bf q}\right)\exp\Big[i{\bf q}\cdot{\bf d}_{\mu}\Big].
\end{equation}
Then the Bragg peaks corresponding to AIAO ordering are located  at ${\bf G}$, provided  the form factor at that reciprocal lattice vector does not vanish. Few examples of the Bragg peaks and values of their  form factors  are $\{{\bf G}=2\pi(2,0,2), f=16/3)\}$,   $\{{\bf G}=2\pi(2,2,0), f=16/3)\}$, $\{{\bf G}=2\pi(2,2,4), f=16/9)\}$, $\{{\bf G}=2\pi(3,3,\pm1), f=128/57)\}$ and $\{{\bf G}=2\pi(1,1,3), f=128/33)\}$. The reason for the vanishing of $f({\bf G})$ at some reciprocal lattice wave-vectors, {\it e.g} $G=(2,0,0), (1,1,1), (3,3,3)$, is the projector 
$({\bf 1}-\frac{{\bf q}{\bf q}}{{\bf q}\cdot{\bf q}})$ in Eq.~\ref{fq} which eliminates the scattering intensity at these  wave-vectors.   

\begin{figure*}[t]
\includegraphics[scale=1.0]{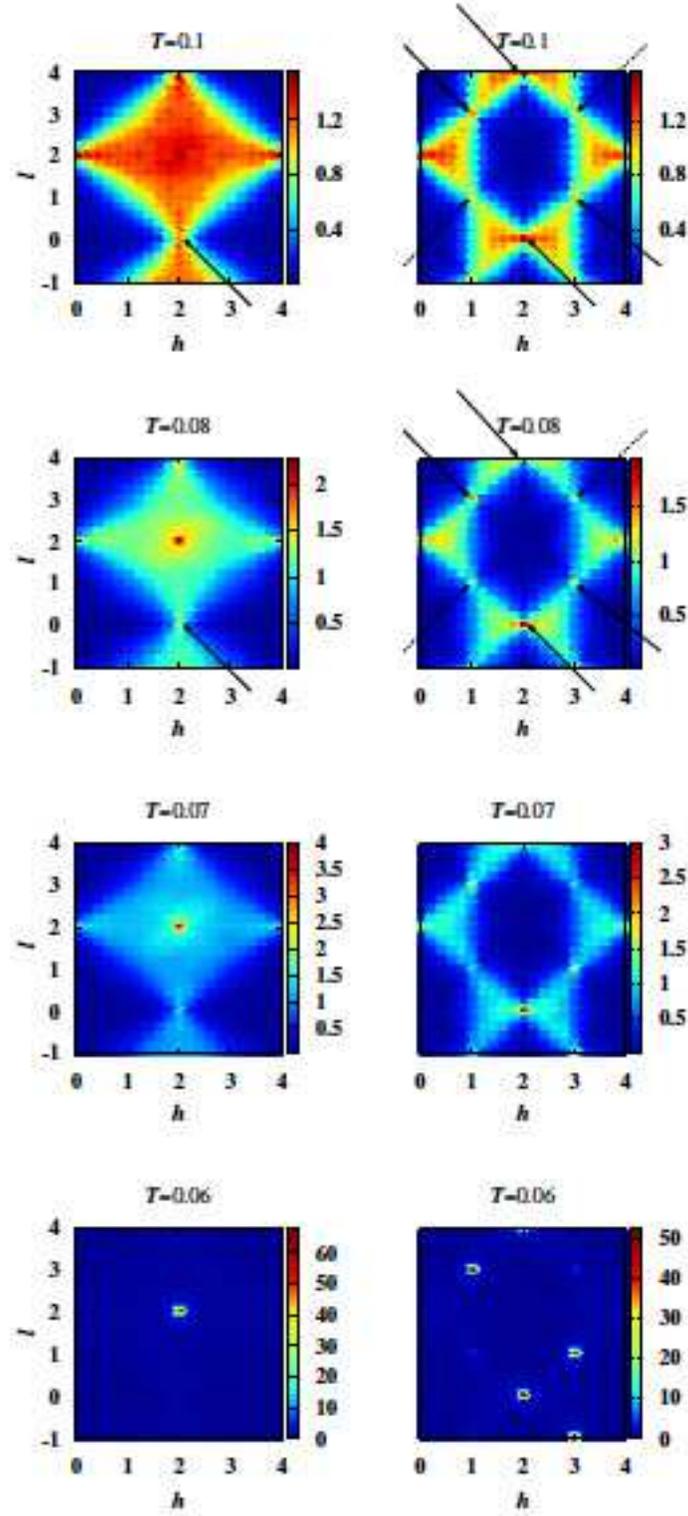}
\caption{Density plot of  neutron structure function $S({\bf q})$ obtained by Monte Carlo simulations  on the pyrochlore lattice of  linear size $L=10$  in the  ({\bf Left column}) [h0l]  and  ({\bf Right column}) [hhl]  planes at the temperatures $T/J_{1}=0.1, 0.08, 0.07, 0.06$. Note that the intensity scale (right color bar) evolves as $T$ approaches $T_c$. The arrows in the top panels, display the location of the pinch-points at $T=0.1, 0.08$}
\label{SF-L10}
\end{figure*}

Fig.~\ref{SF-L10}, illustrates the density plots of $S({\bf q})$ obtained from MC simulations in a lattice of linear size $L=10$, for some temperatures above $T_c$. The thermal averaging has been done over $500$ samples.  This figure clearly represents the pinch-point structures of the nearest-neighbor
Heisenberg pyrochlore antiferromagnet  obtained by Zinkin {\it et al.} \cite{zinkin}.  The location of the pinch-points for $T=0.1$ and $T=0.08$ in [hhl] plane,  shown by arrows in right panel of Fig.~\ref{SF-L10}, are at the wave vectors $(1,1,1), (1,1,3), (2,2,0), (3,3,3), (2,2,4)$. The pinch-points in 
the $[h0l]$ plane are located at  $(2,0,0), (0,0,2), (2,0,4), (4,0,2)$.   

 Close to the transition temperature,  the Bragg peaks corresponding to AIAO ordering begin to grow, which as expected are located at $(2,0,2), (2,2,0), (2,2,4), (1,1,3)$ and $(3,3,\pm1)$ in Fig.~\ref{SF-L10}. Some of these points,  e.g $(2,2,0), (2,2,4),(1,1,3),(3,3,1)$  indicated  by solid arrows in right panels of Fig.~\ref{SF-L10} 
correspond to the pinch-points at $T>T_c$, which flare up in intensity upon cooling down toward $T_c$ and 
which finally form the Bragg peaks at $T<T_c$. This is while the other pinch-points, i.e  $(1,1,1),  (3,3,3)$ indicated  by dashed arrows in the 
right panels of Fig.~\ref{SF-L10}, remain as diffuse peaks upon crossing the transition.  It turns out that these are  reciprocal lattice vectors for which one would expect  Bragg peaks if there were  long range coplanar nematic order.


\end{document}